\documentclass[submission,Phys]{SciPost}

\providecommand{\eprint}[2][]{}
\renewcommand{\eprint}[2][]{%
  \href{https://arxiv.org/abs/#2}{%
    \nolinkurl{https://arxiv.org/abs/#2}}%
}

\hypersetup{
    colorlinks,
    linkcolor={red!50!black},
    citecolor={blue!50!black},
    urlcolor={blue!80!black}
}

\usepackage[bitstream-charter]{mathdesign}
\DeclareSymbolFont{usualmathcal}{OMS}{cmsy}{m}{n}
\DeclareSymbolFontAlphabet{\mathcal}{usualmathcal}

\usepackage{comment}
\usepackage{physics}
\usepackage{mathtools,bm}
\usepackage{amsthm}
\usepackage{xcolor}
\usepackage{tikz}

\newcommand{\WT}{\widetilde{\mathbb T}}
\newcommand{\phd}{\phantom{\dagger}}
\newcommand{\der}{\partial}
\begin{document}

\begin{center}
{\Large\textbf{Arctic Curves and a Gapped Gas Phase in a Two-Band Free-Fermion Chain}}
\end{center}

\begin{center}
Charles Jordan\textsuperscript{1,*},
Dimitri M. Gangardt\textsuperscript{2} and
Alexander G. Abanov\textsuperscript{1}
\end{center}

\begin{center}
{\bf 1} Department of Physics and Astronomy, Stony Brook University,
Stony Brook, NY 11794, USA
\\
{\bf 2} School of Physics and Astronomy, University of Birmingham,
Edgbaston, Birmingham B15 2TT, UK
\\
\textsuperscript{*}\href{mailto:charles.jordan@stonybrook.edu}
{charles.jordan@stonybrook.edu}
\end{center}

\begin{center}
\today
\end{center}

\section*{Abstract}
{\bf
We study the imaginary-time evolution of a domain-wall state constrained to return to itself in a staggered free-fermion chain. The resulting space-time profile separates into frozen, liquid, and gas phases divided by sharp boundaries known as arctic curves. The spectral gap produces an incompressible half-filled gas phase bounded by an inner arctic curve, in addition to the outer frozen--liquid boundary. We determine both arctic curves, the thermodynamic return amplitude, and the complete equal-time correlation kernel. Correlations decay algebraically in the liquid regions and exponentially in the gas, while both arctic boundaries arise as caustics of free quasiparticle trajectories. The return amplitude and correlation kernel are respectively controlled by the determinant and inverse of the same block-Toeplitz operator. We obtain an exact matrix Wiener--Hopf factorization of this operator by reducing the problem to a scalar Riemann--Hilbert problem on an elliptic spectral curve. The factorization also yields the exact thermodynamic return amplitude:
its logarithm consists of a quadratic term with explicit gap dependence
and a bounded periodic theta-function correction. The resulting frozen--liquid--gas structure is a continuous-time free-fermion counterpart of that found in doubly periodic dimer models.
}

\vspace{10pt}
\noindent\rule{\textwidth}{1pt}
\begingroup
\setcounter{tocdepth}{2}
\tableofcontents
\endgroup
\thispagestyle{fancy}
\noindent\rule{\textwidth}{1pt}
\vspace{10pt}

\section{Introduction}
\label{sec:intro}

Limit-shape phenomena arise in many settings, including random tilings,
random partitions and large Young diagrams
\cite{vershik1977asymptotics,Okounkov2001InfiniteWedge,
Borodin2016Lectures}, and stochastic growth and directed polymers
\cite{Corwin2012Kardar}. In the thermodynamic limit, a fluctuating
microscopic system develops an essentially deterministic macroscopic
profile, often separated into regions with qualitatively different
behavior by sharp boundaries known as arctic curves. The canonical
example is the arctic circle of random domino tilings of the Aztec
diamond, which separates four frozen corners from a fluctuating interior
\cite{jockusch1998random}; see
Refs.~\cite{kenyon2009lectures,Gorin2021Lectures,StephanRandomTilings}
for reviews and for
connections between random tilings and free fermions. In the quantum
setting, conditioning a free-fermion state to contain a macroscopically
large empty interval similarly produces an astroid-shaped arctic curve
\cite{Abanov2006Hydrodynamics}.

The exact solution of planar dimer models goes back to Kasteleyn's
Pfaffian formulation \cite{Kasteleyn1961TheStatistics}; their macroscopic
limit shapes are governed by a variational principle
\cite{Cohn2001AVariational}. The uniform Aztec diamond contains only
frozen and liquid regions. A richer phase structure becomes possible when
the dimer weights are periodic. The general theory of weighted bipartite dimer models on
doubly periodic planar graphs associates to the model an algebraic
spectral curve whose amoeba organizes its frozen, liquid, and gas phases
\cite{Kenyon2006Dimers}. One of the simplest finite-domain realizations
is the two-periodic Aztec diamond. In this model, a central gas region,
characterized by bounded height fluctuations and exponentially decaying
correlations, is separated from the surrounding liquid by an inner
arctic curve
\cite{Boutillier2007PatternDensities,chhita2016domino,duits2020two}.
The geometry of the resulting
phase diagram has recently been analyzed for general doubly periodic
Aztec dimer models in Ref.~\cite{BerggrenBorodin2023}. Building on this algebro-geometric viewpoint, recent work has further
developed general variational and integrable frameworks for dimer limit
shapes \cite{Bobenkos2026,bobenko2024dimersmcurves}. Complementary
conformal and PDE methods have established general regularity and
algebraicity results for frozen boundaries \cite{Astala2026}. This
frozen--liquid--gas structure is the direct classical counterpart of
the phase diagram that we find below for a gapped two-band
free-fermion chain. 
Appendix~\ref{app:alternative-scalar-reduction-factor-swapping}
makes the methodological connection explicit by adapting the factor-swapping
construction of periodic dimer models to the present continuous-time symbol.

The connection between random tilings and quantum chains follows from their transfer-matrix description. Taking one direction as Euclidean time maps a tiling to nonintersecting particle trajectories, with fermionic statistics producing determinantal processes and extended correlation kernels \cite{Okounkov2003CorrelationFunction,Eynard1998MatricesCoupled,Borodin2011DeterminantalPoint,StephanRandomTilings}. Conversely, fermionic matrix elements between prescribed boundary states can be interpreted as partition functions of two-dimensional world-line ensembles. A central example is the emptiness formation probability, studied by hydrodynamic, bosonization, and Toeplitz-determinant methods \cite{Abanov2006Hydrodynamics,yeh2022emptiness,Abanov2025EmptinessInstanton,Abanov2002OnTheProbability,Abanov2003EmptinessFormation,Franchini2005Asymptotics}, and more broadly as an extreme event in fermionic number statistics \cite{Marino2016NumberStatistics,Corwin2018CoulombGas,Pallister2025PhaseTransitions}. For domain-wall boundary conditions, a homogeneous free-fermion chain produces an arctic circle and an inhomogeneous critical theory inside it \cite{allegra2016inhomogeneous,bocini2021non}. The same space-time viewpoint relates the Gross--Witten--Wadia transition \cite{gross1980possible,wadia1980n} to the merger of liquid regions separated by a frozen domain \cite{Pallister_2022}. Together, these results show how global fermionic observables can be understood through the geometry of space-time density profiles.

Motivated by this correspondence, we consider the simplest multiband extension of the homogeneous free-fermion problem: a staggered chain with a two-site unit cell and alternating on-site potential \(\pm\Delta\). Its one-particle Hamiltonian is the uniform-hopping, staggered-potential limit of the Rice--Mele chain \cite{Rice1982ElementaryExcitations}. The staggering splits the spectrum into two bands and opens a gap at half filling. We impose the same domain-wall state at both temporal boundaries of a Euclidean slab, with one half of the chain empty and the other completely filled, and ask how the usual arctic-circle picture is modified by the band gap. Preliminary numerical evidence for this model indicated a central half-filled gapped region surrounded by fluctuating and frozen regions \cite{allegra2016inhomogeneous}. Here we establish this phase structure analytically. We derive the boundaries and local correlations of the central region, identify it as a gapped gas phase, and relate its appearance to the compact real component of an elliptic spectral curve. Specifically, the usual outer frozen--liquid boundary survives, while a second arctic curve emerges within the liquid region. This second arctic curve encloses an incompressible phase in which the lower band is filled, the upper band is empty, and the density is pinned to one fermion per unit cell---one half of its maximal value. Correlations in this region decay exponentially, making it the fermionic analogue of the gas phase in periodic dimer models. A common block-Toeplitz structure underlies both the complete transition kernel and the thermodynamic return amplitude. The emergence of the gas phase is not specific to a staggered on-site potential: Appendix~\ref{app:rice-mele-extensions} shows that the leading return free energy and arctic geometry extend to the full Rice--Mele chain, including the Su--Schrieffer--Heeger (SSH) limit.

The main technical result is an exact canonical Wiener--Hopf factorization of the \(2\times2\) matrix propagator. Exploiting its Chebotarev--Khrapkov structure \cite{Chebotarev1956,Khrapkov1971}, we reduce the matrix problem to a scalar Riemann--Hilbert problem on an elliptic spectral curve. We solve this problem by adapting Antipov's Baker--Akhiezer construction \cite{Antipov_2014}, expressing the result in terms of Abelian differentials and genus-one theta functions. The same factorization determines the return amplitude, the local correlation kernel, and the characteristics whose caustics form the two arctic curves.

The paper is organized as follows. Section~\ref{sec:model} defines the
staggered free-fermion chain, the domain-wall return amplitude, and the
Euclidean transition kernel. Section~\ref{sec:main_results} summarizes
the main results and their physical interpretation.
Section~\ref{sec:Toeplitz} expresses the observables in terms of
block-Toeplitz operators and takes the thermodynamic limit.
Section~\ref{sec:WienerHopf} constructs the associated Wiener--Hopf
factorization through a scalar problem on the elliptic spectral curve.
Section~\ref{sec:return amplitude} evaluates the return
amplitude. Section~\ref{sec:saddle-geometry} derives the saddle geometry
and the two arctic curves, while Section~\ref{sec:transition-kernel}
extracts the local kernels, densities, and correlation functions. The appendices contain the finite-rank-source derivation, the
steepest-descent and correlation details, a factorization-flow derivation
of the Wiener--Hopf factors, and a factor-swapping derivation connecting
the calculation to periodic dimers.
Appendix~\ref{app:hydrodynamic-picture} gives a complementary
hydrodynamic interpretation of the characteristic curve and its caustics,
while Appendix~\ref{app:rice-mele-extensions} extends the leading return
free energy and arctic geometry to the full Rice--Mele chain.

\section{The Model}
\label{sec:model}

\begin{figure}[t]
    \centering
    \includegraphics[width=\textwidth]{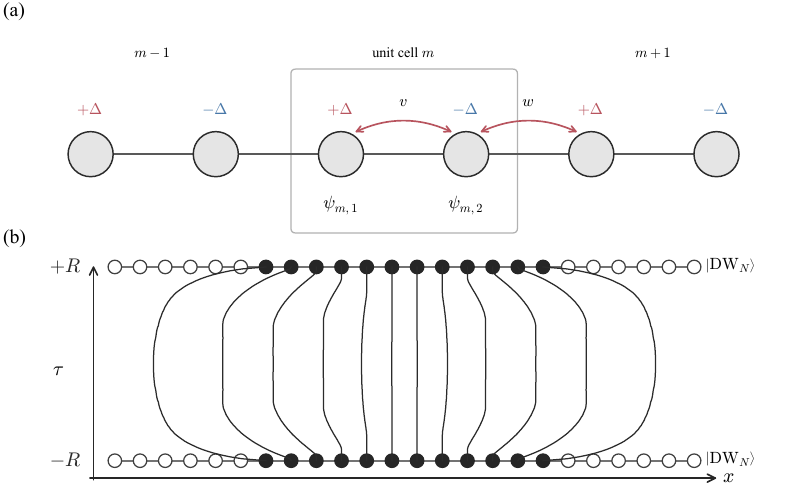}
    \caption{
        Model and Euclidean return geometry.
        \textbf{(a)} Rice-Mele free-fermion chain with a two-site unit cell
        (boxed). The red and blue orbitals have on-site potentials
        \(+\Delta\) and \(-\Delta\), respectively. The nearest-neighbor hopping amplitudes alternate between \(v\) and \(w\) in the general Rice--Mele chain; throughout the main text we specialize to uniform hopping, \(v=w=1\).
        \textbf{(b)} Euclidean slab of height \(2R\) defining the finite-interval return amplitude \(Z_N(R)=\langle \mathrm{DW}_N|e^{-2R\widehat H}|\mathrm{DW}_N\rangle.\) In the boundary state \(|\mathrm{DW}_N\rangle\), both orbitals are occupied in the cells \(m=0,\ldots,N-1\), while all other cells are empty. The figure shows both domain walls bounding the occupied interval.
    }
    \label{fig:model-return-geometry}
\end{figure}

Consider non-interacting fermions hopping on a  one-dimensional chain with alternating chemical potential. We
define unit cells labeled by $m\in \mathbb{Z}$ to contain two lattice sites,
one with chemical potential $\Delta$ and the other with $-\Delta$, see Fig.~\ref{fig:model-return-geometry}a. Fermion annihilation operators $\psi_n$ on
the lattice site $n$ are re-arranged
\begin{align}
\psi_{2m}=\psi_{m,1}, &  & \psi_{2m+1}=\psi_{m,2},
\end{align}
in terms of the operators $\psi_{m,a}$ with integer unit-cell coordinate $m$.
We use Latin letters $m,n,j,\ell$ for cell indices and $a,b,c,d\in\{1,2\}$
for sublattice indices. Band and spectral-sheet labels are denoted by
Greek letters $\sigma,\lambda\in\{-,+\}$.
Sublattices \(1\) and \(2\) carry on-site potentials \(+\Delta\)
and \(-\Delta\), respectively.   The fermionic operators obey canonical anti-commutation relations,
\begin{align*}
    \qty{\psi^{\phd}_{m,a},\psi^\dagger_{n,b}} =\delta_{mn}\delta_{ab}\, .
\end{align*}
The natural two-band generalization of this chain is the Rice--Mele model
\cite{Rice1982ElementaryExcitations},
\begin{align}
\widehat H_{\rm RM}
&=\sum_m\left(
v\,\psi^\dagger_{m,1}\psi^{\phd}_{m,2}
+w\,\psi^\dagger_{m,2}\psi^{\phd}_{m+1,1}
+\text{h.c.}\right)
\nonumber\\
&\quad
+\Delta\sum_m\left(
\psi^\dagger_{m,1}\psi^{\phd}_{m,1}
-\psi^\dagger_{m,2}\psi^{\phd}_{m,2}\right),
\label{eq:rice-mele-hamiltonian}
\end{align}
where \(v,w>0\) are the intracell and intercell hopping amplitudes,
respectively. In the main text we specialize to the uniform-hopping limit
\(v=w=1\), while retaining the staggered on-site potential \(\Delta\).
We denote the resulting many-body Hamiltonian by \(\widehat H\):
\begin{align}
\widehat H
&=\sum_m\left(
\psi^\dagger_{m,1}\psi^{\phd}_{m,2}
+\psi^\dagger_{m,2}\psi^{\phd}_{m+1,1}
+\text{h.c.}\right)
+\Delta\sum_m\left(
\psi^\dagger_{m,1}\psi^{\phd}_{m,1}
-\psi^\dagger_{m,2}\psi^{\phd}_{m,2}\right)
 \nonumber\\
&=\sum_{a,b=1}^{2}\oint\frac{\dd z}{2\pi iz}\,
\psi_a^\dagger(z)\mathbf h_{ab}(z)\psi_b^{\phd}(z).
\end{align}
The extension of the leading return and arctic geometry to arbitrary
\(v,w>0\), including the SSH limit \(\Delta=0\), is given in
Appendix~\ref{app:rice-mele-extensions}.
Here we use the Fourier convention
\begin{equation}
\psi_a(z)=\sum_{m\in\mathbb Z}z^{-m}\psi_{m,a},
\qquad
\psi_{m,a}
=\oint_{|z|=1}\frac{dz}{2\pi iz}\,z^m\psi_a(z).
\label{eq:fourier-convention}
\end{equation}
and the one-particle Hamiltonian
represented by the $2\times2$ Bloch matrix
\begin{equation}
\mathbf h(z)=  
\begin{pmatrix}  
\Delta & 1+z^{-1}\\
1+z & -\Delta  
\end{pmatrix}.
\label{eq:hz}
\end{equation}
We use hats for many-body quadratic operators, such as $\widehat H$, and
boldface for their $2\times2$ one-particle Bloch matrices, such as
$\mathbf h(z)$. Fermion creation and annihilation operators remain
unhatted. Calligraphic letters will denote operators on the one-particle
Hilbert space, while ordinary capital letters will denote finite matrices
and finite-rank sources.

We will be using extensively the complex notation $z=e^{ik}$ for quasimomenta $k$. In terms of these variables the energy spectrum consists of two bands  with dispersion 
\begin{align*}
    \pm\varepsilon(k) = \pm\sqrt{4\cos^2\frac{k}{2} +\Delta^2}  =\pm \sqrt{\qty(z+\kappa)\qty(1+\frac{1}{\kappa z})} = E_{\pm}(z)\, .
\end{align*}
The parameter 
\begin{equation}
 \kappa=
 \frac{1}{2}\left(\Delta^2+2+|\Delta|\sqrt{\Delta^2+4}\right)>1,
 \label{eq:kappa}
\end{equation}
so that \(\kappa+\kappa^{-1}=A=\Delta^2+2\) will be used extensively in what follows. For $\Delta=0$ we have $\kappa=\kappa^{-1}=1$ and the spectrum reduces to the standard cosine dispersion  $\pm 2\cos\frac{k}{2}$ in the reduced Brillouin zone, $-\pi/2<k/2<\pi/2$. 
We write
\[
E_\sigma(e^{ik})=\sigma\varepsilon(k),
\qquad \sigma=-,+,
\]
where \(\sigma=-\) and \(\sigma=+\) label the lower and upper bands,
respectively.

We define the finite domain-wall state by filling cells $0,1,\dots N-1$,
\[
|\mathrm{DW}_N\rangle
=\prod_{m=0}^{N-1}\psi_{m,1}^\dagger\psi_{m,2}^\dagger|0\rangle,
\]
and consider it as  the initial state at Euclidean time $\tau=-R$ and the final state at $\tau=+R$.  We are interested  in the imaginary time return amplitude 
\begin{align}\label{eq:ZR}
Z_N(R)=\langle\mathrm{DW}_N|e^{-2R\widehat H}|\mathrm{DW}_N\rangle,
\end{align}

The return amplitude \eqref{eq:ZR} admits several physical
interpretations. It is the Laplace transform of the energy distribution
in the domain-wall state, with \(2R\) conjugate to energy, and is closely
related to generating functions of quantum work
\cite{Silva2008Work}. In the world-line representation, it is the
partition function of nonintersecting directed polymers with fixed
endpoints, and the density profile describes their macroscopic shape
\cite{Pallister2025PhaseTransitions}. The intermediate densities defined below may also be interpreted as weak
values between states obtained by imaginary-time filtering from the two
temporal boundaries \cite{Aharonov1988WeakValues}; such filtering can be
implemented through measurement protocols
\cite{Mao2023ImaginaryTime}.

 In addition to the return amplitude, we study equal-time fermionic
correlation functions inside the Euclidean slab. To define them, it is
useful to collect the two orbitals in the column vector
\[
\boldsymbol\psi_m
=\begin{pmatrix}\psi_{m,1}\\\psi_{m,2}\end{pmatrix},
\]
so that \(\boldsymbol\psi_m^\dagger\) is a row vector. Since the fermions
are noninteracting, all correlation functions follow from the two-point
kernel by Wick's theorem. We define the equal-time particle kernel as
the \(2\times2\) matrix
\begin{align}\label{eq:corrker}
[\mathbf K_N(m,n;\tau)]_{ab}
=\frac{\langle\mathrm{DW}_N|e^{-(R-\tau)\widehat H}
\psi_{n,b}^{\dagger}\psi^{\phd}_{m,a}e^{-(R+\tau)\widehat H}
|\mathrm{DW}_N\rangle}{Z_N(R)},
\qquad -R\leq\tau\leq R.
\end{align}
Thus $\mathbf K_N(m,n;\tau)$ is the $(m,n)$ block of the one-particle
density operator. Notice that it is not the matrix
$\langle\boldsymbol\psi_m\boldsymbol\psi_n^\dagger\rangle$, which is the
corresponding hole kernel $\delta_{mn}I-\mathbf K_N(m,n;\tau)$.

The average density at site $(m,a)$ at time $\tau$ is then given by  the diagonal element of the above kernel,
\[
\rho_{N,a}(m,\tau)=[\mathbf K_N(m,m;\tau)]_{aa}
\]
and the connected density-density correlation function is obtained via 
 Wick's theorem, \emph{e.g.} for the correlations of total densities in  cells $m>n$ we have 
\begin{equation}
C_{\mathrm{cell},N}(m,n;\tau) 
=
-\operatorname{Tr}\!\left[\mathbf K_N(m,n;\tau)\mathbf K_N(n,m;\tau)\right].
\label{eq:cell-correlation-kernel0}
\end{equation}

We will be ultimately interested in the case when $N\to \infty$, so the boundary states $\ket{\mathrm{DW}_\infty}$ consist of empty sites for $m<0$ and filled ones for $m\geq0$. 
After taking the limit  $N\to \infty$, we consider $R\to\infty$ limit and study correlation functions in terms of the scaled macroscopic variables
\[
x=\frac{m}{R},
\qquad
t=\frac{\tau}{R}\in[-1,1]\, .
\]
The kernel \eqref{eq:corrker}, the average density and the density-density correlation functions
are evaluated explicitly in Section~\ref{sec:transition-kernel}.

\section{Main results}
\label{sec:main_results}

We now summarize the main results, leaving their derivation to the
following sections. Our main results are the frozen--liquid--gas phase
diagram and its two arctic curves, the local correlation kernel and the
phase-dependent correlation asymptotics, the thermodynamic return
amplitude, and the exact Wiener--Hopf factorization underlying all three.

\subsection{Frozen--liquid--gas phase diagram}

The domain-wall return geometry produces five regions in space-time. To
describe them, we denote the local occupation fractions of the lower and
upper bands by $\rho_-(x,t)$ and $\rho_+(x,t)$, respectively. The average
density per lattice site is
\[
\bar\rho(x,t)=\frac{\rho_-(x,t)+\rho_+(x,t)}{2}.
\]
Along the horizontal line $t=0$, one encounters, from left to right, an
empty frozen region, a liquid region, a central gas region, a second
liquid region, and a fully occupied frozen region. Their densities obey
\[
0,\qquad 0<\bar\rho<\frac12,\qquad
\bar\rho=\frac12,\qquad
\frac12<\bar\rho<1,\qquad 1,
\]
respectively. The outer arctic curve separates the liquids from the two
frozen regions, while an inner arctic curve separates the liquids from
the gas.

\begin{figure}[t]
    \centering
    \includegraphics[width=\textwidth]{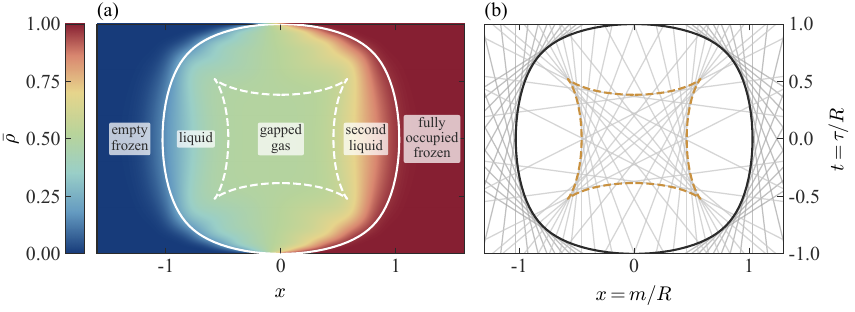}
    \caption{Space-time phase diagram for \(\Delta=2\).
\textbf{(a)}  Finite-\(R\) average density
\(
\bar\rho(m,\tau)
=\frac12\operatorname{Tr}\mathbf K_N(m,m;\tau),
\)
computed from Eq.~\eqref{eq:finite-kernel-operator} at \(R=5\) and shown
in the rescaled coordinates \(x=m/R\) and \(t=\tau/R\). The numerical
calculation uses a 64-cell open chain with \(N=32\) occupied cells in the
boundary state. The solid and dashed white curves are the analytic outer
and inner arctic curves obtained from
Eq.~\eqref{eq:results-arctic-curves} and are superimposed without fitting.
\textbf{(b)} Representative free-quasiparticle trajectories
\(x_\sigma(t,z)=x_{0,\sigma}(z)+t\,v_\sigma(z)\). Their envelopes give the
outer frozen--liquid boundary and the inner liquid-gas boundary
    }
    \label{fig:principal-phase-diagram}
\end{figure}

The inner curve is produced by the spectral gap. For $\Delta=0$, the gap
is closed and the inner boundary disappears. A nonzero $\Delta$ opens a
gap at half filling and pins the gas density to one fermion per unit cell,
or one half of the maximal density. In this region the lower band is
filled and the upper band is empty,
\[
\rho_-^{\mathrm{gas}}=1,
\qquad
\rho_+^{\mathrm{gas}}=0,
\qquad
\bar\rho_{\mathrm{gas}}=\frac12.
\]
Thus the word \emph{gas} does not mean an empty region: the gas is a
translation-invariant, gapped band insulator between two fluctuating
liquid regions. The complete phase diagram is shown in
Fig.~\ref{fig:principal-phase-diagram}.
A central half-filled gapped region in this model was previously observed
numerically in Fig.~15 of Ref.~\cite{allegra2016inhomogeneous}. Here we
identify this region as a gas phase, determine its inner boundary exactly,
and calculate its local densities and correlations.

\subsection{Arctic curves as quasiparticle caustics}

On the hydrodynamic scale, a free quasiparticle with spectral parameter
$z$ and band label $\sigma=\pm$ moves along a straight characteristic,
\[
x_\sigma(t,z)=x_{0,\sigma}(z)+t\,v_\sigma(z),
\qquad
v_\sigma(z)=z\der_zE_\sigma(z).
\]
The group velocity $v_\sigma (z)$ follows directly from the dispersion $E_\sigma(z)$. The nontrivial
quantity is the midpoint position $x_{0,\sigma}(z)$, which contains the
effect of imposing the same domain-wall state at the two temporal
boundaries. Determining this function is one of the main results of our
calculation.

The two branches of the dispersion are conveniently encoded by the
elliptic spectral curve
\begin{equation}
\eta^2(z)=z^2E^2(z)=z(z+\kappa)(z+\kappa^{-1}).
\label{eq:elliptic-curve}
\end{equation}
In the exact solution, the return boundary condition is encoded by a
normalized Abelian differential on this curve. Its normalization
introduces the constant
\begin{equation}
c=-\frac{1}{\kappa}
\frac{\mathrm E(1-\kappa^2)}{\mathrm K(1-\kappa^2)},
\label{eq:results-kappa-c}
\end{equation}
where $\mathrm K$ and $\mathrm E$ are the complete elliptic integrals.
Our exact solution gives
\begin{equation}
x_\sigma(t,z)=x_{0,\sigma}(z)+t\,v_\sigma(z),
\qquad
v_\sigma(z)=z\der_zE_\sigma(z)
=\sigma\frac{z^2-1}{2\eta(z)},
\qquad
x_{0,\sigma}(z)
=\sigma\frac{z^2-2cz+1}{2\eta(z)}.
\label{eq:results-free-flight}
\end{equation}

Evaluation of the local kernel shows that the average density is encoded
in the phase of the relevant saddle point,
\[
z(x,t)=e^{u(x,t)}e^{2\pi i\bar\rho(x,t)},
\qquad
u(x,t)=\log|z(x,t)|.
\]
Here the argument, $2\pi \bar\rho$, is assumed to be  continuous, and is fixed to be
zero in the left frozen region. As one moves from left to right through
the phase diagram, it reaches \(2\pi\), corresponding to $\bar \rho = 1$ in 
the right frozen region.
Solving Eq.~\eqref{eq:results-free-flight} for the corresponding saddle
therefore determines $\bar\rho(x,t)$. In a liquid region, the saddle
equation has a complex-conjugate pair of solutions. Furthermore, at an arctic curve,
the pair reaches the real axis and coalesces. The arctic curves are thus
caustics, or envelopes of the free trajectories, and therefore satisfy
$\partial_zx_\sigma(t,z)=0$.

Solving the characteristic equation together with this caustic condition
gives the two boundaries in parametric form,
\begin{equation}
\boxed{
\begin{aligned}
t(z)
&=-\frac{z^4+2(A+c)z^3-2(A+c)z-1}
{z^4+2Az^3+6z^2+2Az+1},
\\
x(z)
&=\pm
\frac{(1+t(z))z^2-2cz+(1-t(z))}
{2\sqrt{z(z+\kappa)(z+\kappa^{-1})}},
\qquad A=\kappa+\kappa^{-1}.
\end{aligned}}
\label{eq:results-arctic-curves}
\end{equation}
The interval $z\in(0,\infty)$ gives the outer frozen--liquid curve,
whereas $z\in(-\kappa,-\kappa^{-1})$ gives the inner liquid--gas curve.
The curves in the figures are obtained by sampling these two real
intervals in Eq.~\eqref{eq:results-arctic-curves}. A hydrodynamic
derivation of the midpoint data and the caustic condition is given in
Appendix~\ref{app:hydrodynamic-picture}.

\begin{figure}[t]
    \centering
    \includegraphics[width=\textwidth]{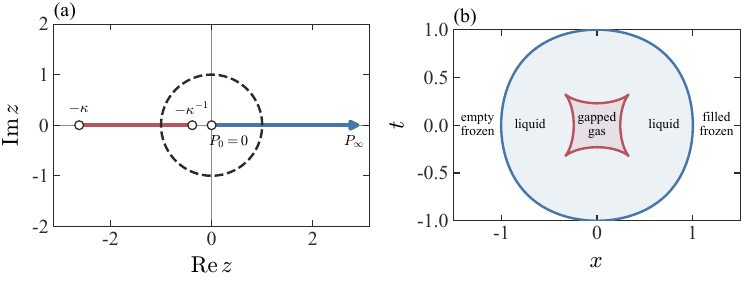}
    \caption{
     Spectral origin of the two arctic boundaries.
\textbf{(a)} Branch-cut representation of the elliptic spectral curve
\(\eta^2=z(z+\kappa)(z+\kappa^{-1})\) in the complex \(z\)-plane. The
cuts are chosen along the positive interval \([0,\infty)\) and the
compact negative interval \([-\kappa,-\kappa^{-1}]\); their lifts form
the two real components of the spectral curve. Both cuts cross the unit
circle \(|z|=1\), shown by the dashed line.
\textbf{(b)} Images of these intervals in the \((x,t)\) plane under the
parametric map in Eq.~\eqref{eq:results-arctic-curves}, obtained from the
envelope condition \(\partial_zx_\sigma(t,z)=0\). The positive interval
gives the outer frozen--liquid boundary, while the compact negative
interval gives the inner liquid--gas boundary.} 
    
    \label{fig:spectral-origin-boundaries}
\end{figure}

\subsection{Local densities and correlations}

The local kernel contains more information than the average density. Its
diagonal entries give the two sublattice densities, $\rho_1$ and $\rho_2$, while its
band-resolved components give the lower- and upper-band occupation
fractions, $\rho_\mp$. These quantities are derived in
Section~\ref{sec:transition-kernel} and displayed in
Fig.~\ref{fig:subdensities}.

Although the total gas density is fixed at one particle per cell, the
staggered potential makes the two sublattice densities unequal:
\begin{equation}
\begin{aligned}
\rho_1^{\mathrm{gas}}
&=\frac12
-\frac{\Delta}{\pi\sqrt{\Delta^2+4}}\,
\mathrm K\!\left(\frac{4}{\Delta^2+4}\right),\\
\rho_2^{\mathrm{gas}}
&=\frac12
+\frac{\Delta}{\pi\sqrt{\Delta^2+4}}\,
\mathrm K\!\left(\frac{4}{\Delta^2+4}\right).
\end{aligned}
\label{eq:results-gas-density}
\end{equation}

The distinction between liquid and gas is especially clear in their
correlations. Consider two cells $m$ and $n$ centered at the macroscopic
point
\[
x=\frac{m+n}{2R},
\qquad
t=\frac{\tau}{R},
\]
and denote their microscopic separation by $r=m-n$. Thus $x$ and $t$ are
rescaled coordinates, whereas $r$ is measured in unit cells. We denote by
$C_{\mathrm{cell}}(r;x,t)$ the corresponding local scaling limit of the
finite-$N$ correlation function defined in
Eq.~\eqref{eq:cell-correlation-kernel0}. The large-distance formulas below
apply for $1\ll|r|\ll R$.

 In a liquid region,
\begin{equation}
C_{\mathrm{cell}}^{\mathrm{liquid}}(r;x,t)
=
-\frac{1}{2\pi^2r^2}
\left[
1-T_F\cos\!\bigl(4\pi r\bar\rho(x,t)\bigr)
\right]
+O(r^{-3}).
\label{eq:results-liquid-correlation}
\end{equation}
Here \(T_F\) is a smooth overlap of the band projectors at the two local
Fermi points. The oscillations with an \(r^{-2}\) envelope are the usual $2k_F$
Friedel oscillations of a one-dimensional Fermi liquid with a sharp
occupation edge.

In the gas the lower band is completely filled and there is no Fermi
edge. Instead,
\begin{equation}
C_{\mathrm{cell}}^{\mathrm{gas}}(r)
=-\frac{\kappa-\kappa^{-1}}{4\pi|r|}
\kappa^{-2|r|}
\left[1+O(|r|^{-1})\right],
\qquad
\xi_{\mathrm{gas}}=\frac{1}{2\log\kappa}.
\label{eq:results-gas-correlation}
\end{equation}
Correlations are therefore algebraic in the liquids and exponentially
clustered in the gas. The derivation and the explicit expression for
$T_F$ are given in Section~\ref{sec:transition-kernel}.

\begin{figure}[t]
    \centering
    \includegraphics[width=\textwidth]{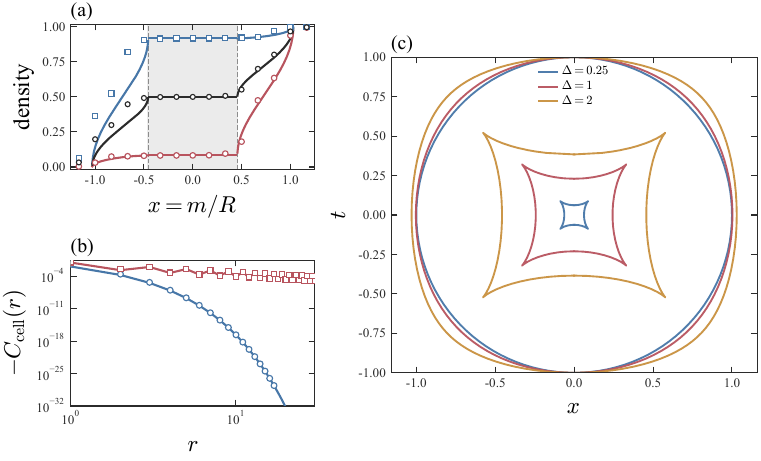}
    \caption{
        Gap dependence and diagnostics of the gas phase.
        \textbf{(a)} Finite-chain sublattice densities at \(t=0\),
        \(\Delta=2\), and \(R=6\) are plotted in red circles for $\rho_{1}$, blue squares for $\rho_{2}$, and black circles for $\bar{\rho}$. Inside the shaded gas region their
        average is \(1/2\) and the individual densities approach
        Eq.~\eqref{eq:results-gas-density}. The analytic results for these densities presented in Section~\ref{sec:transition-kernel} are plotted in solid lines. 
        \textbf{(b)} Connected cell-density correlations for a completely
        filled lower band and for a representative partially filled band.
        The former decay exponentially as in
        Eq.~\eqref{eq:results-gas-correlation}, whereas the latter show the
        \(r^{-2}\) liquid behavior of
        Eq.~\eqref{eq:results-liquid-correlation}. Numerical results are shown for the gapped gas (blue) and a representative liquid state (red), together with the corresponding large-$r$ asymptotic results. 
        \textbf{(c)} Outer frozen--liquid and inner liquid--gas boundaries for several values of \(|\Delta|\). As the gap increases, the inner boundary opens and grows, while the outer boundary changes more weakly. The inner boundary collapses at \(\Delta=0\).
    }
    \label{fig:gap-dependence-diagnostics}
\end{figure}

\subsection{Return amplitude}

The thermodynamic return amplitude is
$Z(R)=\lim_{N\to\infty}Z_N(R)$. Its leading large-$R$ behavior is
\begin{equation}
\boxed{
\log Z(R)
=-4cR^2+O(1)
=\frac{4}{\kappa}
\frac{\mathrm E(1-\kappa^2)}
     {\mathrm K(1-\kappa^2)}R^2+O(1).
}
\label{eq:results-return-asymptotic}
\end{equation}
More precisely, Eq.~\eqref{eq:return-amplitude-exact} gives \(Z(R)\)
exactly as \(e^{-4cR^2}\) times a positive, bounded, and periodic
theta-function quotient. Thus the \(O(1)\) term in
Eq.~\eqref{eq:results-return-asymptotic} is known explicitly.

The finite occupied interval has two domain walls. When \(N\to\infty\) at fixed \(R\), their contributions decouple, up to corrections that are exponentially small in their separation. Reflection combined with particle--hole symmetry relates the two contributions. They have the same leading large-\(R\) free energy, so the coefficient in Eq.~\eqref{eq:results-return-asymptotic} is twice the corresponding single-domain-wall coefficient. In particular, there is no term linear in \(R\). The exact evaluation is
given in Section~\ref{sec:return amplitude}, with details in
Appendix~\ref{app:return-finite-gap}.

\subsection{Exact solution by block Toeplitz and Wiener--Hopf methods}

All results presented above follow from exact representations of the return
amplitude and the correlation kernel in terms of $N\times N$
block-Toeplitz matrices with $2\times2$ blocks, equivalently
$2N\times2N$ matrices. Their symbol is the Euclidean one-particle
propagator
\[
\mathbf W_R(z)=e^{-2R\mathbf h(z)},
\]
where $\mathbf h(z)$ is given by Eq.~\eqref{eq:hz}. The return amplitude is controlled by the determinant of the compressed
propagator, namely the propagator projected onto the occupied subspace.
The transition kernel is controlled by its inverse.

The main technical result is the canonical matrix Wiener--Hopf
factorization
\begin{equation}
\mathbf W_R(z)
=\boldsymbol\Gamma_+(z;R)\boldsymbol\Gamma_-(z;R)^{-1}.
\label{eq:results-WH-schematic}
\end{equation}
Here \(\boldsymbol\Gamma_\pm\), which we call the canonical factors, are
\(2\times2\) matrix functions: \(\boldsymbol\Gamma_+\) and its inverse
are analytic inside the unit circle, while \(\boldsymbol\Gamma_-\) and
its inverse are analytic outside it, including at infinity. A
factorization with these analyticity and invertibility properties is
called canonical. We fix the normalization
\(\boldsymbol\Gamma_-(\infty;R)=I\).
This factorization gives the inverse half-line Toeplitz operator
directly and therefore determines the local kernel and the arctic
geometry. The same factors
also determine the return amplitude. In particular, Widom's differential
formula reduces to
\begin{align}
\partial_R\log Z(R)
={}&-\frac{2}{2\pi i}\oint_{|z|=1}
\operatorname{Tr}\Bigl[
\mathbf h(z)\bigl(
\partial_z\boldsymbol\Gamma_-(z;R)\boldsymbol\Gamma_-(z;R)^{-1}
\nonumber\\[-1mm]
&\hspace{43mm}
-\partial_z\boldsymbol\Gamma_-(z;-R)\boldsymbol\Gamma_-(z;-R)^{-1}
\bigr)\Bigr]\,dz.
\label{eq:results-Widom-differential}
\end{align}
Together with $Z(0)=1$, this identity determines the thermodynamic return
amplitude. Appendix~\ref{app:WH-Toeplitz} explains, using symmetry arguments, why the left and right domain-wall
contributions corresponding to each term in the r.h.s. of Eq.~\eqref{eq:results-Widom-differential} can be written using the same factor
$\boldsymbol\Gamma_-$ evaluated at $R$ and $-R$.

For the staggered chain, the $2\times2$ factorization problem reduces to
a scalar Riemann--Hilbert problem on the elliptic curve
\eqref{eq:elliptic-curve}. Its solution is expressed through normalized
Abelian differentials, a genus-one theta function, and a rank-one
rational correction. Section~\ref{sec:WienerHopf} derives the
factorization \eqref{eq:results-WH-schematic} by the Chebotarev--Khrapkov method. The appendices give two
complementary derivations: Appendix~\ref{app:factorization-flow} obtains
the factors from their isospectral evolution with $R$, while
Appendix~\ref{app:alternative-scalar-reduction-factor-swapping} uses
successive factor swapping and makes contact with periodic dimer models.

\section{From Fermion Observables to Block-Toeplitz Operators}
\label{sec:Toeplitz}

In this section, we reduce the many-body problem to a block-Toeplitz
problem. Since the domain-wall state is a Slater determinant, its return
amplitude is the determinant of the one-particle Euclidean propagator
projected onto the initially occupied subspace, that is, of its
compression to this subspace. The same compressed propagator also
controls local observables: its inverse, dressed by the propagators from
the insertion time to the temporal boundaries, gives the transition
kernel. We first establish these identities at finite \(N\) and then
take \(N\to\infty\), where the determinant of a  finite block-Toeplitz matrix can be expressed in terms of the determinant
of a half-line block-Toeplitz operator, so that the return amplitude becomes its Widom
constant. Thus the problem is reduced to computing the determinant and
inverse of a single block-Toeplitz operator.

\subsection{The compressed propagator and return amplitude}

Using the Fourier convention \eqref{eq:fourier-convention}, the one-particle Hamiltonian is represented by the Bloch matrix \(\mathbf h(z)\) defined in Eq.~\eqref{eq:hz}.
The one-particle Hilbert space has the infinite basis
\(\ket{j,a}\), where \(j\in\mathbb Z\) labels the unit cell and
\(a\in\{1,2\}\) labels the sublattice; we denote this space by \(\mathcal H\).
Let \(\mathcal U(s)\) 
denote the Euclidean propagator on the one-particle
Hilbert space with matrix elements
\begin{align*}
   \bra{m,a} \mathcal U (s) \ket{n,b} \equiv \qty[ \mathcal U(m,n; s)]_{ab} = \oint_{|z|=1}\frac{dz}{2\pi iz}\,z^{m-n}\,\qty[\mathbf U(s; z)]_{ab},
\end{align*}
using the \(2\times2\) symbol
\[
\mathbf U(s;z)=e^{-s\mathbf h(z)}.
\]
We define \(\mathbf W_R(z)=\mathbf U(2R;z)\). The domain-wall return amplitude only
involves its restriction to the \(2N\)-dimensional subspace \(\mathcal H_N\)
generated by \(\ket{j,a}\) with \(0\leq j\leq N-1\) and
\(a\in\{1,2\}\).

We can project the one-particle states onto the orbitals occupied by $\ket{\text{DW}_{N}}$ via the projector
\[
\mathcal P_{N}=\sum_{j=0}^{N-1}\sum_{a=1}^{2}\ket{j,a}\bra{j,a}
\]
which projects $\mathcal{H}$ onto $\mathcal{H}_{N}$. The boundary state $\ket{\mathrm{DW}_N}$
is the Slater determinant obtained by filling an orthonormal basis of
$\mathcal H_N$. A standard property of Slater determinants is that a many-body matrix element of a number-conserving quadratic evolution equals the determinant of the corresponding one-particle overlap matrix \(A_N(R)\), given by
\[
A_N(R)=
\mathcal P_N\mathcal U(2R)\mathcal P_N\big|_{\mathcal H_N}.
\]
Thus the overlap
matrix is the compression of $\mathcal U(2R)$ to $\mathcal H_N$.
It follows that
\begin{equation}
Z_N(R)=\det_{\mathcal H_N}A_N(R)\,.
\label{eq:zdeta}    
\end{equation}

The one-particle propagator is translation invariant in the unit-cell
coordinate, so its matrix elements depend only on the difference of the
cell indices. Thus \(A_N\) is the block-Toeplitz matrix
\begin{equation*}
 A_N(R)=T_N[\mathbf W_R]
 =\bigl((\mathbf W_R)_{j-\ell}\bigr)_{j,\ell=0}^{N-1},
 \qquad
 (\mathbf W_R)_r=\oint_{|z|=1}\frac{dz}{2\pi iz}\,z^r\mathbf W_R(z),
\end{equation*}
where \((\mathbf W_R)_r\) are the \(2\times2\) Fourier coefficients of the symbol.
We denote finite block-Toeplitz matrices by $T_N[\cdot]$ and the
corresponding half-line operators, introduced below, by $\mathsf T[\cdot]$.
We therefore obtain the return amplitude as the following  block-Toeplitz determinant,
\[
Z_N(R)=\det T_N[\mathbf W_R].
\]

\subsection{Finite-rank sources and the transition kernel}

Local observables are controlled by the inverse $A_N^{-1}$ of the same compressed propagator.  To show this, let $F$ be a finite-rank operator  on the one-particle Hilbert space, and let  \(\widehat F\) denote the corresponding fermionic bilinear.  We  define a generating function
\[
\mathcal Z_N[F;\tau]
=
\langle\mathrm{DW}_N|
e^{-(R-\tau)\widehat H}e^{\widehat F}e^{-(R+\tau)\widehat H}
|\mathrm{DW}_N\rangle \,, \qquad      \mathcal Z_N[0;\tau] = Z_N(R)
\]
The Slater-determinant identity gives the exact generating formula
\begin{equation}
\frac{\mathcal Z_N[F;\tau]}{Z_N(R)}=
\det\left[I+(e^F-I)\mathcal C_N(\tau)\right]\,,
\label{eq:finite-source-generating}
\end{equation}
where
\[
\mathcal C_N(\tau)=
\mathcal U(R+\tau)\,\mathcal P_NA_N^{-1}\mathcal P_N\,\mathcal U(R-\tau).
\]
The short derivation, including the reduction of the source determinant to the support of $F$, is given in Appendix~\ref{app:finite-rank-sources}.

Choosing the rank-one source $F=\lambda|n,b\rangle\langle m,a|$ and differentiating at $\lambda=0$ gives the transition kernel. In block notation,
\begin{equation}
\mathbf K_N(m,n;\tau)=
\langle m|\mathcal U(R+\tau)\mathcal P_NA_N^{-1}\mathcal P_N\mathcal U(R-\tau)|n\rangle,
\label{eq:finite-kernel-operator}
\end{equation}
where the matrix element on the right denotes a $2\times2$ block in
the sublattice space.

The transition kernel is therefore obtained by propagating from the
insertion time to the initially occupied subspace, applying the inverse
of the compressed propagator, and propagating back out. Its diagonal is
the local density defined in Section~\ref{sec:model}.

Higher derivatives of \eqref{eq:finite-source-generating} produce
multipoint correlators and reproduce the determinantal structure implied
by Wick's theorem \cite{Eynard1998MatricesCoupled,
Borodin2011DeterminantalPoint}. The Gaussian transition state is thus
encoded in the single operator $\mathcal C_N(\tau)$.

The return amplitude, Eq.~\eqref{eq:zdeta},   depends on the determinant of $A_N$, while the transition kernel, Eq.~\eqref{eq:finite-kernel-operator}, depends on its inverse.
As we show in Appendix \ref{app:finite-rank-sources}
they  can be seen as two aspects of the same Toeplitz problem.
Both  quantities follow from  the elementary variational identity
\begin{equation}
\delta\log\det A_N=
\operatorname{Tr}\left(A_N^{-1}\delta A_N\right).
\label{eq:deltalogdet}
\end{equation}
When the variation $\delta A_N$ involves change of the global parameter $R$, it gives the differential formula for the return amplitude, while when $\delta A_N$ is a localized finite-rank perturbation, the identity \eqref{eq:deltalogdet} produces the transition kernel. 

\subsection{The thermodynamic limit}

We now take the thermodynamic limit $N\to \infty$ at fixed $R$.  
The large-$N$ behavior of the return amplitude is  governed by the block
Szeg\H{o}--Widom theorem
\cite{widom1974asymptotic,widom1976asymptotic,
basor2017asymptotics,Bottcher1990Analysis},
\[
Z_N(R) = \det T_N[\mathbf W_R]=
G[\mathbf W_R]^N E[\mathbf W_R][1+o(1)].
\]
Here, $G[\mathbf W_{R}]$ is the geometric mean
\[
G[\mathbf W_R]=
\exp\left[
\frac{1}{2\pi}
\int_{-\pi}^{\pi}
\log\det \mathbf W_R(e^{ik})\,dk
\right].
\]
In the present model, $\operatorname{tr}\mathbf h(z)=0$, and therefore
\[
\det \mathbf W_R(z)=
e^{-2R\operatorname{tr}\mathbf h(z)}=1.
\]
Since the extensive Szeg\H{o} contribution vanishes, \(G[\mathbf W_R]=1\), the thermodynamic return amplitude is the Szeg\H{o}--Widom constant
\cite{widom1976asymptotic,Bottcher1990Analysis},
\[
Z(R)
:=
\lim_{N\to\infty}Z_N(R)
=
E[\mathbf W_R]
=
\det\!\left(
\mathsf T[\mathbf W_R]\mathsf T[\mathbf W_R^{-1}]
\right).
\]
Here $\mathbf W_R^{-1}(z)=\mathbf W_{-R}(z)$. In this limit
the finite matrix $T_N[\mathbf W_R]$ is replaced by the half-line block-Toeplitz
operator $\mathsf T[\mathbf W_R]$, acting on two-component one-particle states with
cell index $m\geq0$. 
Physically, the finite state \(\ket{\mathrm{DW}_N}\) has two domain walls, one at each end of the occupied interval. In the limit \(N\to\infty\) at fixed \(R\), the two walls separate and their contributions to \(\log Z_N(R)\) become independent, with corrections that vanish exponentially with their separation. The resulting Widom constant contains both edge contributions. Reflection combined with particle--hole symmetry relates them, and their leading large-\(R\) free energies are equal.

For \(z=e^{ik}\), the Bloch Hamiltonian \(\mathbf h(e^{ik})\) is Hermitian, and hence
\[
\mathbf W_R(e^{ik})=e^{-2R\mathbf h(e^{ik})}
\]
is Hermitian and positive definite. Its finite block-Toeplitz compressions are therefore positive definite, so the finite-\(N\) return amplitude
\(Z_N(R)=\det T_N[\mathbf W_R]\) is positive and well defined. The same
uniform positivity implies that the half-line Toeplitz operator
\(\mathsf T[\mathbf W_R]\) is invertible, as required for the transition
kernel. We have therefore reduced both observables to the same block-Toeplitz problem: the return amplitude is given by the Szeg\H{o}--Widom constant defined above, while the transition kernel is obtained from \(\mathsf T[\mathbf W_R]^{-1}\). The remaining task is to construct the canonical Wiener--Hopf factors of the matrix symbol \(\mathbf W_R(z)\).

\section{Exact Matrix Wiener--Hopf Factorization}
\label{sec:WienerHopf}

We now construct the canonical Wiener--Hopf factors of the symbol
$\mathbf W_R(z)=e^{-2R\mathbf h(z)}$. The main idea is simple. The staggered-chain
symbol belongs to the Chebotarev--Khrapkov class, so its matrix
eigenvalue problem defines a two-sheeted elliptic curve. On that curve
the matrix factorization becomes a scalar Riemann--Hilbert problem. Its
solution has three ingredients: Abelian exponentials carrying the
$R$-dependent singularities, a theta quotient restoring
single-valuedness, and a rank-one rational factor removing the auxiliary
divisor. We keep these ingredients visible here; the Toeplitz identities
and the analytic checks are given in Appendices~\ref{app:WH-Toeplitz}
and~\ref{app:WH-analytic-details}.

\subsection{Canonical factors and their two uses}

Since \(\mathbf W_R(e^{ik})\) is Hermitian positive definite, it
admits the canonical Wiener--Hopf factorization
\cite{Wiener1957ThePrediction,ClanceyGohberg1981,Bottcher1990Analysis},
\begin{equation}
\mathbf W_R(z)=\boldsymbol\Gamma_+(z;R)\boldsymbol\Gamma_-(z;R)^{-1}.
\label{eq:WH-factorization}
\end{equation}
We use the analyticity conditions and normalization stated after
Eq.~\eqref{eq:results-WH-schematic}.
This factorization gives the inverse half-line Toeplitz operator and hence the transition kernel. As shown in Appendix~\ref{app:WH-Toeplitz}, the generating function of its matrix elements is
\begin{equation}
\sum_{\ell,j\geq0}w^{-\ell}z^j
\bigl(\mathsf T[\mathbf W_R]^{-1}\bigr)_{\ell b,ja}
=\left[
\frac{w}{w-z}\boldsymbol\Gamma_-(w;R)\boldsymbol\Gamma_+(z;R)^{-1}
\right]_{ba},
\qquad |z|<|w|.
\label{eq:WH-inverse-generating}
\end{equation}
The same Wiener--Hopf factors also determine the return amplitude. The standard Widom formula has
one contribution from each end of a long Toeplitz section. Reflection
combined with particle--hole symmetry maps the right-end contribution to
the left-end one with $R\mapsto-R$. For the present symbol using this fact and Eq.~\eqref{eq:deltalogdet}  gives the
one-factor formula
\begin{align}
\partial_R\log Z(R)
={}&-\frac{2}{2\pi i}\oint_{|z|=1}
\operatorname{Tr}\Bigl[
\mathbf h(z)\bigl(
\partial_z\boldsymbol\Gamma_-(z;R)\boldsymbol\Gamma_-(z;R)^{-1}
\nonumber\\[-1mm]
&\hspace{43mm}
-\partial_z\boldsymbol\Gamma_-(z;-R)\boldsymbol\Gamma_-(z;-R)^{-1}
\bigr)\Bigr]\,dz.
\label{eq:WH-Widom-differential}
\end{align}
Thus the same factors \(\boldsymbol\Gamma_\pm\) determine both
the transition kernel and the return amplitude. The standard two-ordering
formula and its reduction to Eq.~\eqref{eq:WH-Widom-differential}, together
with the exact double-contour formula for the transition kernel, are
collected in Appendix~\ref{app:WH-Toeplitz}.

\subsection{Reduction to the spectral curve}

\begin{figure}
    \centering
    \includegraphics[width=0.65\linewidth]{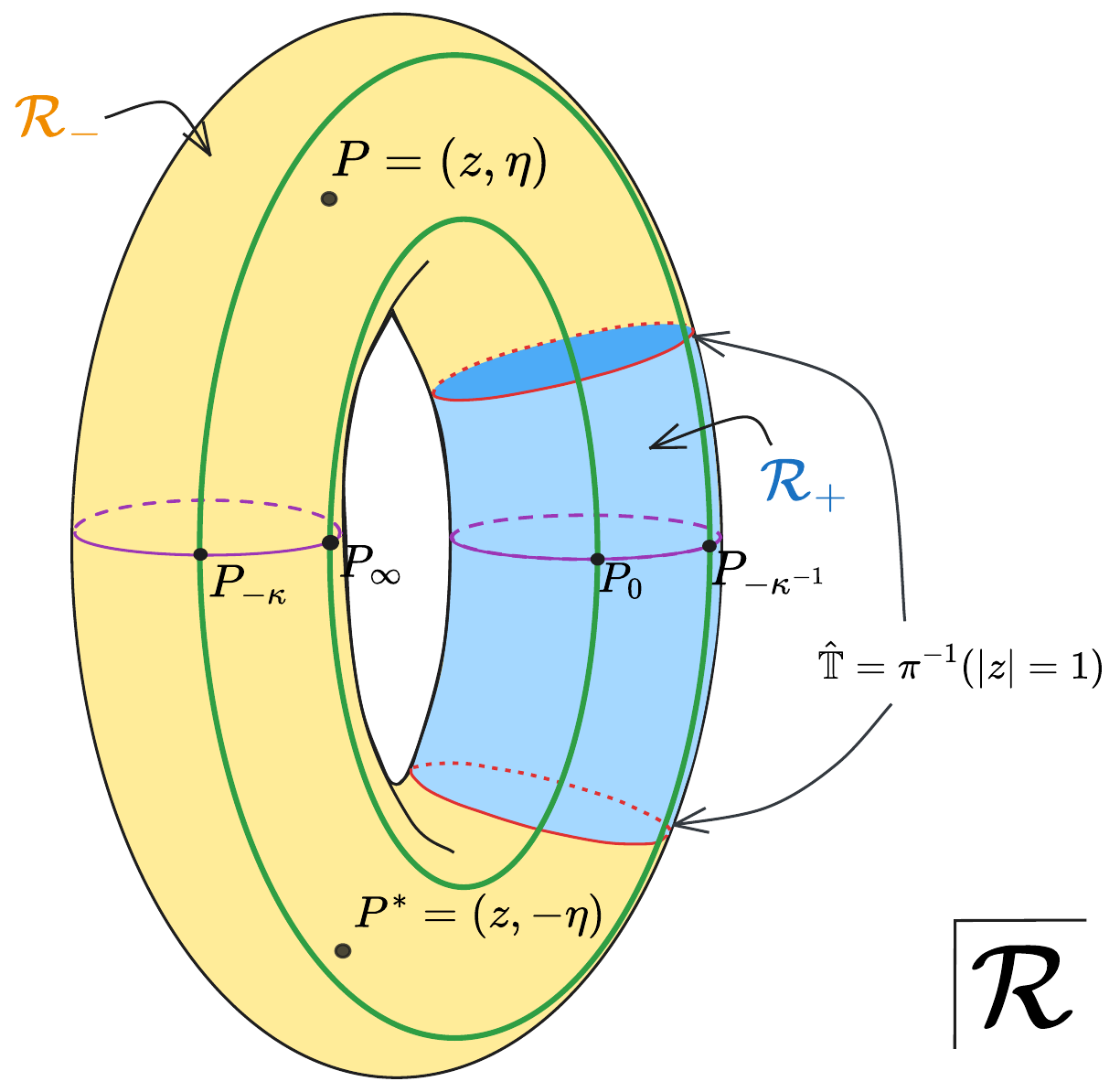}
    \caption{Schematic representation of the elliptic spectral curve $\eta^2(z)=z(z+\kappa)(z+\kappa^{-1})$ which defines a two-dimensional surface in four dimensional space $(z,\eta)$. Each point $P=(z,\eta)$ on the surface has a conjugate point $P^*=(z,-\eta)$, except for  ramification (branch) points $P=P^*$ for which $\eta=0$ or $\infty$.
    The lift $\WT=\pi^{-1} (|z|=1)$ of the Wiener--Hopf contour $|z|=1$ separates two domains, \(\mathcal R_+\) and \(\mathcal R_-\), corresponding to \(|z|<1\) and \(|z|>1\), respectively.  The green lines are $a$-cycles surrounding branch cuts between pairs of ramification (branch)  points $P_{-\kappa}=(-\kappa,0)$, $P_{-\kappa^{-1}}=(-\kappa^{-1},0)$ and $P_{0}=(0,0)$, $P_{\infty}=(\infty,\infty)$, while purple lines are $b$-cycles. }
    \label{fig:elliptic_curve}
\end{figure}

The Chebotarev--Khrapkov structure of the symbol becomes explicit after
removing the negative power of \(z\) from the Bloch Hamiltonian. Define
\begin{equation}
\mathbf Q(z)=z\mathbf h(z)=
\begin{pmatrix}
\Delta z&1+z\\
z(1+z)&-\Delta z
\end{pmatrix}.
\label{eq:Q-ChebKh}
\end{equation}
This matrix is traceless and satisfies the scalar-square identity
\[
\mathbf Q(z)^2=f(z)I,
\qquad
f(z)=z(z+\kappa)(z+\kappa^{-1}),
\]
where
\[
\kappa=\frac12\left(A+\sqrt{A^2-4}\right)>1,
\qquad
A=\Delta^2+2.
\]
Consequently, functions of \(\mathbf Q(z)\), including the symbol
\(\mathbf W_R(z)\), reduce to linear combinations of \(I\) and
\(\mathbf Q(z)\). This places \(\mathbf W_R(z)\) in the
Chebotarev--Khrapkov class
\cite{Chebotarev1956,Khrapkov1971}.
The scalar polynomial \(f(z)\) defines the two-sheeted spectral curve
\begin{equation}
\mathcal R:\qquad \eta^2=f(z).
\label{eq:spectral-curve-WH}
\end{equation}
For $\Delta\neq0$ it has genus one. A point is denoted by
$P=(z,\eta)$, and $P^*=(z,-\eta)$ is the point on the opposite sheet, see Fig.~\ref{fig:elliptic_curve}. Since \(\mathbf Q(z)=z\mathbf h(z)\) and
\(\mathbf Q(z)^2=f(z)I\), the two eigenvalues of \(\mathbf h(z)\)
are \(\pm\sqrt{f(z)}/z\). On the spectral curve, they are represented
by the single meromorphic function
\[
E(P)=\frac{\eta(P)}{z(P)},
\qquad E(P^*)=-E(P).
\]
Thus the two sheets encode the two energy bands.
The corresponding rank-one spectral
projectors are
\[
\mathbf Y(P)=\frac12\left(I+\frac{\mathbf Q(z)}{\eta}\right),
\qquad
\mathbf Y(P^*)=I-\mathbf Y(P),
\]
and hence
\begin{equation}
\mathbf W_R(z)=e^{-2RE(P)}\mathbf Y(P)+e^{-2RE(P^*)}\mathbf Y(P^*).
\label{eq:WR-spectral-decomposition}
\end{equation}
Although the two terms live on different sheets, their sum is invariant
under $P\leftrightarrow P^*$ and is therefore a single-valued matrix on
the $z$-plane.

Let $\WT=\pi^{-1}(|z|=1)$ be the lift of the unit circle
to $\mathcal R$. The regions over $|z|<1$ and $|z|>1$ will be denoted by
$\mathcal R_+$ and $\mathcal R_-$, respectively. The branch point
$P_0=(0,0)$ lies in the closure of $\mathcal R_+$, while $P_\infty$ lies
in the closure of $\mathcal R_-$ as shown in Fig.~\ref{fig:elliptic_curve}.

To solve the dual factorization, we first construct preliminary matrix
factors from scalar functions on the two sheets:
\[
\widetilde{\boldsymbol\Gamma}_\pm(z;R)
=
\chi_\pm(P;R)\mathbf Y(P)
+
\chi_\pm(P^*;R)\mathbf Y(P^*).
\]
Since \(\mathbf Y(P)\) and \(\mathbf Y(P^*)\) are complementary
orthogonal projectors,
\[
\widetilde{\boldsymbol\Gamma}_+(z;R)
\widetilde{\boldsymbol\Gamma}_-(z;R)^{-1}
=
\frac{\chi_+(P;R)}{\chi_-(P;R)}\mathbf Y(P)
+
\frac{\chi_+(P^*;R)}{\chi_-(P^*;R)}\mathbf Y(P^*).
\]
Comparing this expression with the spectral decomposition
\eqref{eq:WR-spectral-decomposition}, we find that the matrix
factorization \eqref{eq:WH-factorization} is equivalent to the scalar
jump
\begin{equation}
\chi_+(P;R)=e^{-2RE(P)}\chi_-(P;R),
\qquad P\in\WT.
\label{eq:scalar-WH-jump}
\end{equation}
Thus the noncommutative \(2\times2\) problem on the unit circle reduces
to a scalar problem on \(\mathcal R\).

\subsection{Solution of the scalar problem}

The scalar problem \eqref{eq:scalar-WH-jump} consists in finding functions
\(\chi_+\) and \(\chi_-\), analytic in the Wiener-Hopf domains \(\mathcal R_+\) and
\(\mathcal R_-\), respectively, whose ratio across \(\WT\) is
\(e^{-2RE(P)}\). Following Ref.~\cite{Antipov_2014}, we construct them in two steps: first we reproduce
the exponential jump, and then we remove the monodromy of the
resulting functions around the cycles of the spectral curve.

Since the genus-one surface \(\mathcal R\) has the topology of a torus,
it is convenient to describe its points by a coordinate on a complex
torus. Choose canonical \(a\)- and \(b\)-cycles on \(\mathcal R\), with
the \(a\)-cycle surrounding the compact cut
\([-\kappa,-\kappa^{-1}]\), as shown in
Fig.~\ref{fig:elliptic_curve}. The normalized holomorphic differential is
\[
d\omega=\frac{1}{\mathcal A}\frac{dz}{\eta},
\qquad
\mathcal A=\oint_a\frac{dz}{\eta},
\qquad
B=\oint_b d\omega,
\]
so that \(\oint_a d\omega=1\) and \(\operatorname{Im}B>0\). Its integral
defines the Abel coordinate
\[
u(P)=\int_{P_{\mathrm{ref}}}^{P}d\omega
\qquad
\operatorname{mod}\bigl(\mathbb Z+B\mathbb Z\bigr),
\]
where \(P_{\mathrm{ref}}\) is a fixed reference point. When the integration
path winds around the \(a\)- or \(b\)-cycle, \(u(P)\) changes by \(1\) or
\(B\), respectively. Analytic continuation on \(\mathcal R\) is therefore
represented by translations on the complex torus. This allows us to use
the quasiperiodicity of the genus-one theta function to cancel the
monodromy encountered below \cite{Fay1973ThetaFunctions}.

The energy \(E(P)\) has simple poles at
\(P_0\in\mathcal R_+\) and \(P_\infty\in\mathcal R_-\). Indeed, in the local coordinates \(u_0=z^{1/2}\) and \(u_\infty=z^{-1/2}\), respectively, \(E(P)\) behaves as \(u_0^{-1}\) and \(u_\infty^{-1}\), up to the choice of sign of the local coordinate. Its
differential therefore has double poles with vanishing residues at these
points. To separate the two singularities between the Wiener--Hopf
domains, we introduce normalized differentials of the second kind,
that is, meromorphic differentials with vanishing residues:
\begin{equation}
d\Omega_\infty=-\frac{z-c}{\eta}\,dz,
\qquad
d\Omega_0=\frac{1-cz}{z\eta}\,dz.
\label{eq:second-kind-WH}
\end{equation}
Here \(d\Omega_\infty\) has its only (double) pole at \(P_\infty\), while \(d\Omega_0\) has its only (double) pole at \(P_0\).
Their \(a\)-periods vanish if we choose
\begin{equation}
c=
\frac{\displaystyle\oint_a z\,dz/\eta}
     {\displaystyle\oint_a dz/\eta}
=
-\frac{1}{\kappa}
\frac{\mathrm E(1-\kappa^2)}{\mathrm K(1-\kappa^2)}.
\label{eq:c-WH}
\end{equation}
The corresponding Abelian integrals can be normalized so that
\[
\Omega_\infty(P)+\Omega_0(P)=-2E(P).
\]
Consequently,
\[
e^{R\Omega_\infty(P)}
=
e^{-2RE(P)}e^{-R\Omega_0(P)}.
\]
Thus \(e^{R\Omega_\infty}\) and \(e^{-R\Omega_0}\) have the required
jump. The first is regular in \(\mathcal R_+\), since its
singularity lies at \(P_\infty\); the second is regular in
\(\mathcal R_-\), since its singularity lies at \(P_0\).

These exponentials need not yet be single-valued. The vanishing
\(a\)-periods remove the monodromy around the \(a\)-cycle, but
continuation around the \(b\)-cycle generally multiplies them by a
constant. Define
\[
V=\frac{1}{2\pi i}\oint_b d\Omega_\infty,
\qquad
K_B=\frac{1+B}{2}.
\]
Since \(d\Omega_\infty+d\Omega_0=-2\,dE\), both exponentials acquire
the same multiplier \(e^{2\pi iRV}\) around the \(b\)-cycle.
We cancel this factor using the genus-one theta function
\[
\theta(u|B)=
\sum_{n\in\mathbb Z}e^{\pi i n^2B+2\pi inu}.
\]
Choose a fixed auxiliary point \(P_D\) on \(\mathcal R\), away from the
jump contour and the branch points, and define
\begin{equation}
q(P;R)=
\frac{\theta(u(P)+K_B-u(P_D)+RV\,|\,B)}
     {\theta(u(P)+K_B-u(P_D)\,|\,B)}.
\label{eq:theta-quotient-WH}
\end{equation}
Under \(u\mapsto u+1\), the quotient  $q(P;R)$ is unchanged, while under
\(u\mapsto u+B\) it acquires the multiplier \(e^{-2\pi iRV}\).
Thus 
\begin{equation}
\widetilde\chi_+(P;R)=e^{R\Omega_\infty(P)}q(P;R),
\qquad
\widetilde\chi_-(P;R)=e^{-R\Omega_0(P)}q(P;R)
\label{eq:scalar-WH-solution}
\end{equation}
are now single-valued meromorphic functions in their respective
Wiener--Hopf domains. Because the same quotient multiplies both
exponentials, they still satisfy
Eq.~\eqref{eq:scalar-WH-jump}. 


\subsection{From the scalar solution to the matrix factors}

The theta function has a simple zero at
\(K_B\) modulo the period lattice. Therefore, the denominator of
\(q(P;R)\) vanishes at the fixed point \(P_D\), while its numerator
vanishes at a point \(P_Z(R)\) determined by
\[
u(P_Z(R))\equiv u(P_D)-RV
\pmod{\mathbb Z+B\mathbb Z}.
\]
For generic \(R\), these points are distinct and lie away from the branch
points, so \(q(P;R)\) has a simple pole at \(P_D\) and a simple zero at
\(P_Z(R)\). These singularities are auxiliary features of the scalar
construction and must be removed when constructing the canonical matrix
factors.

For each \(z\) away from the branch points, its two preimages
\(P\) and \(P^*\) on the spectral curve correspond to the two spectral
projectors \(\mathbf Y(P)\) and \(\mathbf Y(P^*)\). Recombining the scalar
solutions gives
\begin{equation}
\widetilde{\boldsymbol\Gamma}_\pm(z;R)
=
\widetilde\chi_\pm(P;R)\mathbf Y(P)
+
\widetilde\chi_\pm(P^*;R)\mathbf Y(P^*).
\label{eq:preliminary-matrix-WH}
\end{equation}
Exchanging \(P\) and \(P^*\) leaves this expression unchanged,
so it defines a single-valued matrix on the \(z\)-plane.
The projector identities ensure that these matrices satisfy the
required matrix jump.

The pole at \(P_D\) affects only one spectral component and
therefore produces a rank-one matrix residue at
\(z_D=\pi(P_D)\). Similarly, the zero at \(P_Z(R)\) produces a
one-dimensional kernel at \(z_Z(R)=\pi(P_Z(R))\). Let \(v_Z(R)\)
span this right kernel, and let \(\ell_D^\top\) denote the row
direction of the residue at \(z_D\).

Both defects are removed by right multiplication with the
elementary rational matrix
\begin{equation}
\mathbf C(z;R)=I+
\frac{z_Z(R)-z_D}{z-z_Z(R)}
\frac{v_Z(R)\ell_D^\top}{\ell_D^\top v_Z(R)}.
\label{eq:rational-correction-WH}
\end{equation}
At \(z_D\), this matrix annihilates the row direction of the
residue. Its pole at \(z_Z\) is cancelled by the zero of
\(\widetilde{\boldsymbol\Gamma}_\pm\) in the direction \(v_Z\).
The same rational correction is applied to both preliminary factors, so
their Wiener--Hopf ratio is unchanged. Since
\(\mathbf C(\infty;R)=I\), the normalization at infinity gives
\begin{equation}
\boxed{
\boldsymbol\Gamma_\pm(z;R)
=
\widetilde{\boldsymbol\Gamma}_\pm(z;R)\mathbf C(z;R)
\widetilde{\boldsymbol\Gamma}_-(\infty;R)^{-1}.
}
\label{eq:exact-dual-WH-factors}
\end{equation}
By construction, \(\boldsymbol\Gamma_-(\infty;R)=I\), and the normalized
factors satisfy
\[
\mathbf W_R(z)
=
\boldsymbol\Gamma_+(z;R)\boldsymbol\Gamma_-(z;R)^{-1}.
\]
Equation~\eqref{eq:exact-dual-WH-factors}, together with Eqs.~\eqref{eq:scalar-WH-solution}--\eqref{eq:rational-correction-WH}, gives the explicit solution of the matrix Wiener--Hopf problem.
The analyticity and cancellation
checks for generic divisor positions are given in
Appendix~\ref{app:WH-analytic-details}. At exceptional values of \(R\) for which the auxiliary pole and zero coincide or reach a branch point, the normalized factors are obtained by continuity from generic \(R\).

\section{The Return Amplitude}
\label{sec:return amplitude}

We now evaluate the thermodynamic return amplitude using the exact
Wiener--Hopf factors constructed above. The calculation is considerably
shorter than the construction of the factors themselves. The Abelian
exponentials produce the leading quadratic term in \(\log Z(R)\), while
the remaining finite-gap contribution reduces to a theta-function
quotient. This gives an exact expression for \(Z(R)\), from which its
large-\(R\) behavior follows immediately: the correction to the
quadratic term is bounded and periodic, and in particular there is no
term linear in \(R\). The direct extraction of the required coefficient
from the matrix factors is given in
Appendix~\ref{app:return-finite-gap}.

Writing
\[
d\boldsymbol\Gamma_\pm
=(\partial_z\boldsymbol\Gamma_\pm)\,dz
\]
for differentiation with respect to the spectral variable, 
Eq.~\eqref{eq:WH-Widom-differential} becomes
\begin{align}
\partial_R\log Z(R)
={}&-\frac{2}{2\pi i}\oint_{|z|=1}
\operatorname{Tr}\left[
\mathbf h(z)\left(
d\boldsymbol\Gamma_-(R)\boldsymbol\Gamma_-(R)^{-1}
-d\boldsymbol\Gamma_-(-R)\boldsymbol\Gamma_-(-R)^{-1}
\right)\right].
\label{eq:return-Widom-short}
\end{align}
The first term is the contribution of the domain wall at the left end of
the occupied interval. The second is the contribution of the right domain
wall, mapped to the left one by reflection and particle--hole symmetry.
This separation becomes exact in the limit \(N\to\infty\) at fixed \(R\); the
condition $N>R$ alone would not give a strict separation because
imaginary-time propagation has exponentially small tails. The two walls
have the same leading free energy, although their bounded finite-gap
corrections need not agree term by term.

 To evaluate Eq.~\eqref{eq:return-Widom-short}, expand the normalized exterior factor at infinity as
\[
\boldsymbol\Gamma_-(z;R)
=
I+\frac{\boldsymbol\Gamma_1(R)}{z}+O(z^{-2}),
\qquad
g(R)=[\boldsymbol\Gamma_1(R)]_{12}.
\]
Since the only growing entry of \(\mathbf h(z)\) is
\(\mathbf h_{21}(z)=z+1\), evaluation of the contour integral at infinity gives
\begin{equation}
\partial_R\log Z(R)
=
2\bigl[g(R)-g(-R)\bigr].
\label{eq:return-g-reduction}
\end{equation}

Appendix~\ref{app:return-finite-gap} evaluates this coefficient directly from the exact Wiener--Hopf factors. The result is
\begin{equation}
g(R)
=
-2cR+\frac{V}{2}
\left[
\frac{\theta'(D+RV\,|\,B)}{\theta(D+RV\,|\,B)}
-
\frac{\theta'(D\,|\,B)}{\theta(D\,|\,B)}
\right],
\qquad
D\equiv\pm\frac14
\pmod{\mathbb Z+B\mathbb Z}.
\label{eq:return-g-exact}
\end{equation}
Substitution into Eq.~\eqref{eq:return-g-reduction} shows that the
\(R\)-independent term cancels and that the remaining theta contribution
is a total derivative. Using \(Z(0)=1\) and the theta duplication identity,
we obtain
\begin{equation}
\boxed{
Z(R)
=
e^{-4cR^2}
\frac{\theta(2RV\,|\,2B)}
     {\theta(0\,|\,2B)}.
}
\label{eq:return-amplitude-exact}
\end{equation}

\begin{figure}[tbp]
    \centering
    \includegraphics[width=0.85\linewidth]
    {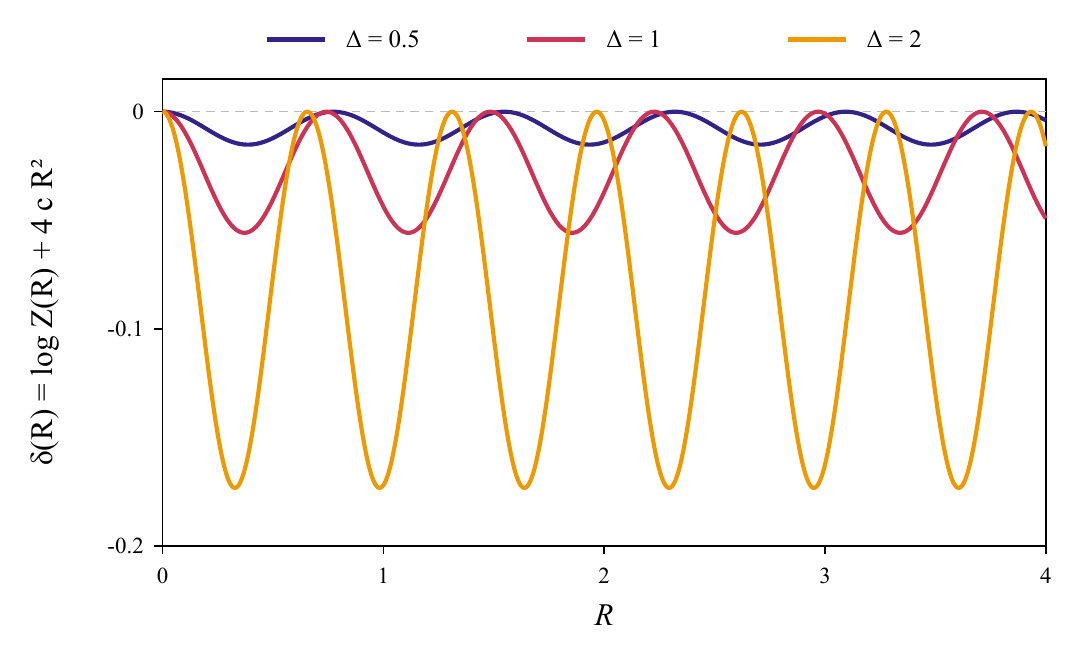}
    \caption{
        Finite-gap correction
        \(\delta(R)=\log Z(R)+4cR^2\), obtained from the exact return
        amplitude \eqref{eq:return-amplitude-exact}, for three values of the
        spectral gap. The correction is bounded and periodic, with period
        \(1/(2|V|)\). Its magnitude increases with the gap for the values
        shown. The dashed line marks \(\delta(R)=0\).
    }
    \label{fig:return-amplitude-remainder}
\end{figure}

For the real cycles used here, $B$ is purely imaginary while $V$ is real. The Jacobi product formula shows that
\(\theta(y\,|\,2B)\) is positive for real \(y\). It is also continuous
and periodic, and therefore bounded above and away from zero. Hence the
theta quotient in Eq.~\eqref{eq:return-amplitude-exact} gives a bounded
periodic correction:
\begin{equation}
\boxed{
\log Z(R)
=
\frac{4}{\kappa}
\frac{\mathrm E(1-\kappa^2)}
     {\mathrm K(1-\kappa^2)}R^2+O(1).
}
\label{eq:return-free-energy}
\end{equation}
The bounded finite-gap correction appearing in
Eq.~\eqref{eq:return-free-energy} is extracted from Eq.~\eqref{eq:return-amplitude-exact} and illustrated in
Fig.~\ref{fig:return-amplitude-remainder}.

Equivalently,
\[
\lim_{R\to\infty}\frac{\log Z(R)}{R^2}
=
-4c
=
\frac{4}{\kappa}
\frac{\mathrm E(1-\kappa^2)}
     {\mathrm K(1-\kappa^2)}.
\]

The exact result is manifestly even in \(R\). This agrees with the
reflection and particle--hole symmetry
\[
\boldsymbol\sigma_y\mathbf h(z^{-1})\boldsymbol\sigma_y
=
-\mathbf h(z),
\]
which implies \(Z_N(-R)=Z_N(R)\) for every finite \(N\). This symmetry is
a useful check on Eq.~\eqref{eq:return-amplitude-exact}, but is not needed
to exclude a term linear in \(R\).

\section{Saddle Geometry and Arctic Curves}
\label{sec:saddle-geometry}

We now derive the saddle geometry and arctic curves announced in
Section~\ref{sec:main_results}. Starting from the exact Wiener--Hopf
factors, we obtain the large-\(R\) saddle equation, identify its real
and complex solutions in the three phases, and derive both arctic curves
from the coalescence of saddles. The local correlation kernel associated
with this geometry is discussed in
Section~\ref{sec:transition-kernel}. A complementary derivation in terms
of free-fermion hydrodynamics is given in
Appendix~\ref{app:hydrodynamic-picture}.

\subsection{Macroscopic scaling and saddle action}

We first take the thermodynamic limit \(N\to\infty\) at fixed \(R\) and denote the resulting kernel by
\begin{equation}
\mathbf K(m,n;\tau)
:=
\lim_{N\to\infty}\mathbf K_N(m,n;\tau).
\label{eq:thermodynamic-kernel}
\end{equation}
The exact kernel has the double-contour representation
\begin{align}
\mathbf K(m,n;\tau)
={}&
\oint_{|z|<|w|}
\frac{dw}{2\pi iw}\frac{dz}{2\pi iz}\,
w^m z^{-n}\frac{w}{w-z}
\nonumber\\[-1mm]
&\hspace{9mm}\times
\mathbf U(R+\tau;w)\boldsymbol\Gamma_-(w;R)
\boldsymbol\Gamma_-(z;R)^{-1}\mathbf U(-(R+\tau);z).
\label{eq:K1}
\end{align}
Here the integration contours are positively oriented nested loops
around the origin, with the \(w\)-contour enclosing the \(z\)-contour.
This representation follows from the inverse Toeplitz generating
function \eqref{eq:WH-inverse-generating}; its derivation is given in
Appendix~\ref{app:WH-Toeplitz}.

We then consider its large-\(R\) behavior near a fixed macroscopic point \((x,t)\). Specifically, we take
\[
m=M+m_0,\qquad n=M+n_0,\qquad \tau=tR,
\qquad \frac{M}{R}\longrightarrow x,
\]
where the integer cell offsets $m_0,n_0$ and $r=m_0-n_0$ remain fixed as $R\to\infty$.
Substituting the exact Wiener--Hopf factors into
Eq.~\eqref{eq:K1} and resolving the resulting matrices into their two
spectral components gives a double integral on the spectral curve,
\begin{align}
\mathbf K(M+m_0,M+n_0;tR)
={}&
\int_{\Gamma_P}\frac{dw}{2\pi iw}
\int_{\Gamma_Q}\frac{dz}{2\pi iz}\,
\frac{w}{w-z}\,w^{m_0}z^{-n_0}
\nonumber\\
&\qquad\times
e^{R[S(P;x_R,t)-S(Q;x_R,t)]}
\mathbf A(P,Q). 
\label{eq:spectral-double-kernel}
\end{align}
Here \(P=(w,\eta(w))\), \(Q=(z,\eta(z))\), and \(x_R=M/R\).
The contours \(\Gamma_P\) and \(\Gamma_Q\) are the full lifts of
the \(w\)- and \(z\)-contours to the spectral curve, including both
sheets and inheriting their orientations. Their projections retain
the original nesting, with the \(w\)-contour enclosing the
\(z\)-contour.
The matrix prefactor $\mathbf A$ contains the spectral projectors, theta
quotients, and rational corrections, but no additional exponential in
$R$. The action in the thermodynamic limit $R\to \infty$ is
\begin{equation}
S(P;x,t)=x\log z-\Omega_0(P)-(1+t)E(P).
\label{eq:saddle-action}
\end{equation}
The contours and the passage from the two sheet-indexed integrals to
Eq.~\eqref{eq:spectral-double-kernel} are described in
Appendix~\ref{app:steepest-descent-topology}.

The saddle condition \(dS=0\) gives the sheet-resolved equation
\begin{equation}
p(z;t)=x\eta(z),
\qquad
p(z;t)=\frac{1+t}{2}z^2-cz+\frac{1-t}{2}.
\label{eq:unsquared-saddle}
\end{equation}
Unlike its squared form below, this equation determines both \(z\) and
the sheet of the spectral curve. Squaring Eq.~\eqref{eq:unsquared-saddle} eliminates the sheet label and
gives the quartic
\begin{equation}
D(z;x,t)=p(z;t)^2-x^2f(z)=0,
\qquad
f(z)=z^3+Az^2+z.
\label{eq:saddle-quartic}
\end{equation}
Since $dS$ has two double poles on a genus-one curve, it has four zeros
counting multiplicity; Eq.~\eqref{eq:saddle-quartic} therefore accounts
for all saddles. For \(x\neq0\), a root \(z_j\) lifts to
\[
P_j=\left(z_j,\frac{p(z_j;t)}{x}\right),
\]
with the sheet fixed by Eq.~\eqref{eq:unsquared-saddle}. The case
\(x=0\), where this expression is singular, is treated separately
below.

\subsection{Free characteristics and phase structure}

The saddle equation has a simple free-particle interpretation. Using the expressions for \(\Omega_0\) and \(E\), the saddle equation
reproduces the free-flight form stated in
Eq.~\eqref{eq:results-free-flight},
\begin{equation}
x_\sigma(t,z)=x_{0,\sigma}(z)+t\,v_\sigma(z),   
\label{eq:saddle-free-flight}
\end{equation}
where
\[
x_{0,\sigma}(z)
=\sigma\frac{z^2-2cz+1}{2\eta(z)},
\qquad
v_\sigma(z)
=\sigma\frac{z^2-1}{2\eta(z)}.
\]
Thus $z$ is conserved and the quasiparticles move along straight
characteristics. The nontrivial information about the domain-wall
boundary state is contained in the midpoint positions
$x_{0,\sigma}(z)$.

The topology of the saddle set changes across the phase diagram:
\begin{align}
\text{gas:}\quad&
\text{four real roots on }(-\kappa,-\kappa^{-1}),\nonumber\\
\text{liquid:}\quad&
\text{two roots on }(-\kappa,-\kappa^{-1})
\text{ and one complex-conjugate pair},\nonumber\\
\text{frozen:}\quad&
\text{two roots on }(-\kappa,-\kappa^{-1})
\text{ and two roots on }(0,\infty).
\label{eq:saddle-root-phases}
\end{align}
\begin{figure}[t]
\centering
\includegraphics[width=\linewidth]{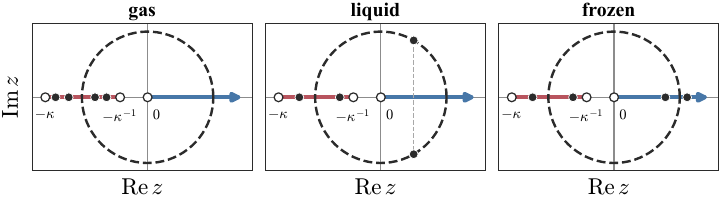}
\caption{ Schematic saddle configurations in the three phases for \(t=0\). The thick
negative interval denotes the compact real component
\((-\kappa,-\kappa^{-1})\) of the spectral curve. In the gas phase, all four
saddles project to this interval. In the liquid phase, two remain on the compact
component while the other two form a complex-conjugate pair on the unit circle \(\lvert z\rvert =1\). Conversely, for \(t\neq0\), this conjugate pair moves along a circle of non-unit radius; and furthermore, for sufficiently large or small radius, this circle no longer intersects the compact real component of the spectral curve, resulting in an absence of the gas phase along the line of a constant $t$. In any case, this pair lies on the positive real axis in a frozen
region. The liquid--gas and
frozen--liquid arctic curves occur when the corresponding pair of saddles
coalesces. The root positions are schematic.}
\label{fig:saddle-configurations}
\end{figure}
These changes in the saddle configuration, illustrated in
Fig.~\ref{fig:saddle-configurations}, determine the arctic boundaries
discussed next. Their relation to the occupied Fermi contours and local
densities is developed in Section~\ref{sec:transition-kernel}, with
contour details given in
Appendix~\ref{app:steepest-descent-topology}.

\subsection{Arctic curves as caustics}

An arctic boundary is reached when two saddles coalesce. This
double-critical-point mechanism is common in exact constructions of
arctic curves \cite{di2014arctic,colomo_arctic_2016}. Equivalently,
the projection of the characteristic surface to the $(x,t)$ plane
becomes singular. Both the outer frozen--liquid boundary and the inner
liquid--gas boundary are therefore components of the same double-root
locus,
\begin{equation}
D(z;x,t)=0,
\qquad
\partial_zD(z;x,t)=0.
\label{eq:arctic-double-root}
\end{equation}
Solving these two equations gives
\begin{align}
t(z)
&=-\frac{(z^2-1)\bigl[z^2+2(A+c)z+1\bigr]}
{z^4+2Az^3+6z^2+2Az+1},
\label{eq:arctic-t-section7}\\
x(z)
&=\frac{2\eta(z)\bigl[2z-c(z^2+1)\bigr]}
{z^4+2Az^3+6z^2+2Az+1}.
\label{eq:arctic-x-section7}
\end{align}
The range $z\in(0,\infty)$ gives the outer frozen--liquid curve, whereas
$z\in(-\kappa,-\kappa^{-1})$ gives the inner liquid--gas curve. The two
signs of $\eta$ generate the two spatially reflected branches.

At $x=0$, the squared equation
\eqref{eq:saddle-quartic} counts both sheets at once, so the intersections
with the symmetry axis must instead be obtained from
$p(z;t)=0$. This minor qualification is important near the endpoints
of the two curves. When $\Delta\to0$, $\kappa\to1$ and the compact
interval collapses; correspondingly, the inner caustic disappears.

\section{Local Kernels and Correlations}
\label{sec:transition-kernel}

We now derive the local kernels, densities, and correlation asymptotics
announced in Section~\ref{sec:main_results}. The local fermionic state is
selected by the saddle geometry and the Cauchy pole in
Eq.~\eqref{eq:spectral-double-kernel}. During steepest descent, the
$P$- and $Q$-contours are deformed in opposite directions. When their
relative nesting changes, they cross the pole $w=z$, and its residue
gives the leading local kernel. The remaining isolated saddle
contributions are smaller by $O(R^{-1})$.

\subsection{The local Fermi contour}

We define \(\mathbf K(r;x,t)\) as the \(R\to\infty\) limit of \(\mathbf K(m,n;\tau)\) with \((m+n)/(2R)\to x\), \(\tau/R\to t\), and \(m-n=r\) fixed.

Let $\gamma(x,t)$ be the oriented contour swept out when the two
steepest-descent contours cross. Away from an arctic boundary,
\begin{equation}
\mathbf K(r;x,t)
=
\int_{\gamma(x,t)}
\frac{dz(P)}{2\pi iz(P)}\,z(P)^r\mathbf Y(P).
\label{eq:local-kernel-contour}
\end{equation}
Corrections to this local limit are \(O(R^{-1})\) away from an arctic boundary.
Thus the large-scale geometry enters the microscopic state only through
the homology class and endpoints of $\gamma$. The contour deformation
leading to Eq.~\eqref{eq:local-kernel-contour} is detailed in
Appendix~\ref{app:steepest-descent-topology}.

Write the complex saddle pair in a liquid region as
$z_\pm=\varrho e^{\pm ik_F}$.
Let \(P_\sigma(k)\) denote the continuous lift of
\(z=\varrho e^{ik}\) associated with the lower \((\sigma=-)\)
or upper \((\sigma=+)\) band.
On the left liquid branch,
$\gamma$ is an occupied arc on the lower-energy sheet:
\begin{equation}
\mathbf K^{\mathrm{left\,liq}}(r;x,t)
=
\int_{-k_F}^{k_F}\frac{dk}{2\pi}\,
\varrho^r e^{ikr}
\mathbf Y(P_-(k)).
\label{eq:left-liquid-kernel}
\end{equation}
The continuous lift \(P_-(k)\) is fixed at the endpoint by
the sheet-resolved saddle equation \eqref{eq:unsquared-saddle}.
\[
\eta(P_-(k_F))
=\frac{p(\varrho e^{ik_F};t)}{x}.
\]
In the gas region the arc closes into the full lower-band cycle,
\begin{equation}
\mathbf K^{\mathrm{gas}}(r)
=
\int_{-\pi}^{\pi}\frac{dk}{2\pi}\,
e^{ikr}\mathbf Y(P_-(k)).
\label{eq:gas-kernel-contour}
\end{equation}
 In the right liquid, the lower band is completely filled and the
upper band is partially filled. It is convenient to describe this
state as a completely filled system with an unoccupied arc removed
from the upper band:
\begin{equation}
\mathbf K^{\mathrm{right\,liq}}(r;x,t)
=\delta_{r0}I
-\int_{-k_F}^{k_F}\frac{dk}{2\pi}\,
\varrho^r e^{ikr}
\mathbf Y(P_+(k)).
\label{eq:right-liquid-kernel}
\end{equation}
Together with the two frozen phases, the result is summarized by
\begin{equation}
\mathbf K(r;x,t)=
\begin{cases}
0, & \text{left frozen},\\
\mathbf K^{\mathrm{left\,liq}}(r;x,t), & \text{left liquid},\\
\mathbf K^{\mathrm{gas}}(r), & \text{gas},\\
\mathbf K^{\mathrm{right\,liq}}(r;x,t), & \text{right liquid},\\
\delta_{r0}I, & \text{right frozen}.
\end{cases}
\label{eq:piecewise-local-kernel}
\end{equation}
 For \(\sigma=-,+\), let
\[
E_\sigma(k):=E(P_\sigma(k)).
\]
The projector along either liquid arc is then
\[
\mathbf Y(P_\sigma(k))
=
\frac12\left(
I+\frac{\mathbf h(\varrho e^{ik})}{E_\sigma(k)}
\right),
\qquad
E_\sigma(k)^2
=
\Delta^2+2+\varrho e^{ik}+\varrho^{-1}e^{-ik}.
\]
The branch of the square root is fixed by the continuous lift
\(P_\sigma(k)\). On the unit circle,
\[
E_\sigma(k)=\sigma\varepsilon(k).
\]
Thus the familiar projectors
\(\tfrac12[I+\sigma\mathbf h(e^{ik})/\varepsilon(k)]\) apply only when
\(\varrho=1\).

\subsection{Densities in the five regions}

The sublattice densities are the diagonal entries of
\(\mathbf K(0;x,t)\). Taking the trace gives the total density per unit
cell,
\begin{equation}
\rho_{\mathrm{cell}}(x,t)
=\operatorname{Tr}\mathbf K(0;x,t)
=
\begin{cases}
0, & \text{left frozen},\\
k_F/\pi, & \text{left liquid},\\
1, & \text{gas},\\
2-k_F/\pi, & \text{right liquid},\\
2, & \text{right frozen}.
\end{cases}
\label{eq:piecewise-density}
\end{equation}

The total density can also be resolved into contributions from the two bands. Let \(\gamma_\sigma^{\mathrm{occ}}(x,t)\) denote the occupied part of the local Fermi contour on the sheet \(\sigma=-,+\), with a completely filled band represented by the corresponding full cycle. We define the local band occupation fractions by
\begin{equation}
\rho_\sigma(x,t)
:=
\int_{\gamma_\sigma^{\mathrm{occ}}(x,t)}
\frac{dz(P)}{2\pi i z(P)},
\qquad \sigma=-,+.
\label{eq:band-density-definition}
\end{equation}
Since each spectral projector has unit trace,
\(\operatorname{Tr}\mathbf Y(P)=1\), this integral is the trace of the
contribution of band \(\sigma\) to \(\mathbf K(0;x,t)\). Thus
\(\rho_\sigma=0\) for an empty band and \(\rho_\sigma=1\) for a
completely filled band. These occupation fractions should not, in
general, be identified with the eigenvalues of the \(2\times2\) matrix
\(\mathbf K(0;x,t)\).

\begin{figure}[t]
    \centering
    \includegraphics[width=\linewidth]{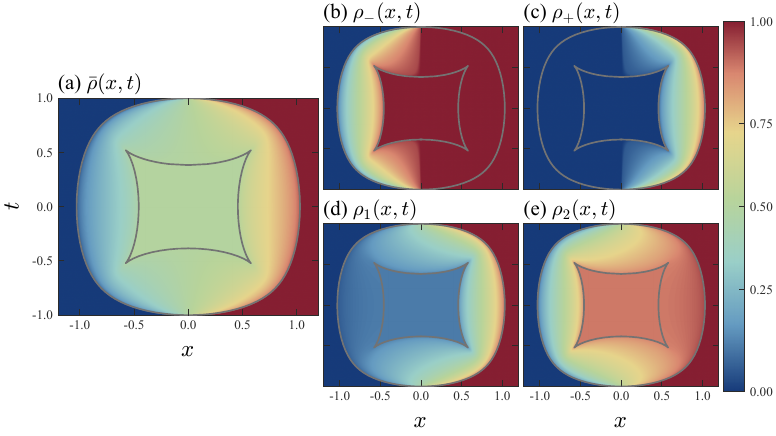}
    \caption{
        Local density observables for \(\Delta=2\).
        \textbf{(a)} The average density per lattice site,
        \(\bar\rho(x,t)=\tfrac12\operatorname{Tr}\mathbf K(0;x,t)\),
        interpolates from the empty frozen region on the left to the fully
        occupied frozen region on the right and is pinned to \(1/2\) in
        the gas.
        \textbf{(b,c)} The lower- and upper-band occupation fractions \(\rho_-(x,t)\) and \(\rho_+(x,t)\), respectively, defined in Eq.~\eqref{eq:band-density-definition}. In the gas, \(\rho_-=1\) and \(\rho_+=0\).
        \textbf{(d,e)} The sublattice densities \(\rho_1(x,t)\) and
        \(\rho_2(x,t)\), given by the diagonal entries of
        \(\mathbf K(0;x,t)\) in the sublattice basis. Their sum gives the
        total density per unit cell, while their difference reflects the
        staggered potential. The curves mark the outer frozen--liquid and
        inner liquid--gas boundaries. 
    }
    \label{fig:subdensities}
\end{figure}

The occupied lower-band arc in the left liquid, the filled lower band in the gas, and the complement of the upper-band hole arc in the right liquid then give
\begin{equation}
\begin{aligned}
\rho_-(x,t)&=
\begin{cases}
0, & \text{left frozen},\\
k_F/\pi\, ,\phantom{+1} & \text{left liquid},\\
1, & \text{gas, right liquid, or right frozen},
\end{cases}\\[1mm]
\rho_+(x,t)&=
\begin{cases}
0, & \text{left frozen, left liquid, or gas},\\
1-k_F/\pi, & \text{right liquid},\\
1, & \text{right frozen}.
\end{cases}
\end{aligned}
\label{eq:piecewise-band-densities}
\end{equation}
Thus
\[
\rho_{\mathrm{cell}}(x,t)=\rho_-(x,t)+\rho_+(x,t),
\qquad
\bar\rho(x,t)=\frac{\rho_-(x,t)+\rho_+(x,t)}{2}.
\]
On the other hand, the sublattice densities \(\rho_1(x,t)\) and \(\rho_2(x,t)\) are given by the 
diagonal entries of \(\mathbf K(0;x,t)\) in the sublattice basis.
The band-resolved and sublattice-resolved observables are displayed in
Fig.~\ref{fig:subdensities}. In the gas, the lower-band occupation is
unity and the upper-band occupation vanishes, while the sublattice
densities are spatially constant but unequal and are given by
Eq.~\eqref{eq:results-gas-density}. Explicit component formulas for the
liquid regions are collected in
Appendix~\ref{app:kernel-correlation-evaluation}.

\subsection{Algebraic and exponential correlations}

For two distinct cells at separation $r$, Wick's theorem gives
\begin{equation}
C_{\mathrm{cell}}(r;x,t)
=-\operatorname{Tr}\!\left[\mathbf K(r;x,t)\mathbf K(-r;x,t)\right].
\label{eq:cell-correlation-kernel}
\end{equation}
In a liquid region, the sharp endpoints of the occupied or hole arc
produce the usual Fermi-edge asymptotic,
\begin{equation}
C_{\mathrm{cell}}^{\mathrm{liquid}}(r;x,t)
=-\frac{1}{2\pi^2r^2}
\left[1-T_F\cos(2rk_F)\right]
+O(r^{-3}),
\label{eq:liquid-correlation-section8}
\end{equation}
where
\[
T_F
=\operatorname{Tr}\!\left[
\mathbf Y(P_F)\mathbf Y(\overline{P_F})
\right].
\]
Here \(P_F\) is the endpoint of the local Fermi arc above
\(z_F=\varrho e^{ik_F}\). More explicitly,
\(P_F=P_-(k_F)\) on the occupied lower-band arc in the
left liquid, and \(P_F=P_+(k_F)\) on the upper-band
hole arc in the right liquid. The other endpoint is its complex
conjugate \(\overline{P_F}\), lying above
\(\overline{z_F}=\varrho e^{-ik_F}\).
Using Eq.~\eqref{eq:piecewise-density}, one has
\[
k_F=\pi-2\pi\left|\bar\rho(x,t)-\frac12\right|.
\]
Since \(r\) is an integer,
\(\cos(2rk_F)=\cos(4\pi r\bar\rho)\), in agreement with the form
stated in Section~\ref{sec:main_results}.

By contrast, in the gas phase  the contour $\gamma$ is closed and there is no Fermi edge. Its large-distance correlation function
is controlled by the nearest branch point of the gapped dispersion:
\begin{equation}
C_{\mathrm{cell}}^{\mathrm{gas}}(r)
=-\frac{\kappa-\kappa^{-1}}{4\pi|r|}
\kappa^{-2|r|}
\left[1+O(|r|^{-1})\right].
\label{eq:gas-correlation-section8}
\end{equation}
Thus
\begin{equation}
\xi_{\mathrm{gas}}
=\frac{1}{2\log\kappa}
=\frac{1}{4\operatorname{arsinh}(|\Delta|/2)}.
\label{eq:gas-correlation-length-section8}
\end{equation}
The endpoint calculation in the liquid and the branch-point calculation
in the gas are given in
Appendix~\ref{app:kernel-correlation-evaluation}.

The order of limits is important. The local-kernel and density formulas above are obtained by first taking
\(R\to\infty\) at fixed \(r\). The correlation
asymptotics then take $|r|\to\infty$ inside a bulk phase. They are not
uniform near an arctic boundary, where a separate edge scaling is
required. In a joint bulk limit one should require $|r|=o(R)$.

\section{Discussion and conclusions}
\label{sec:discussion}

The main physical effect found here is that opening a gap in the
one-particle spectrum creates a new macroscopic region in the Euclidean
space-time profile. In the homogeneous chain the domain-wall geometry
contains frozen and liquid regions. In the staggered chain the gap at half filling creates a gas island
inside the liquid region and pins its density to one fermion per unit
cell. This phase is incompressible but not empty: it is a locally
translation-invariant band insulator. The inner
arctic curve is its boundary with the liquid and may therefore be viewed as
the real-space manifestation of the spectral gap. The different decay of
correlations on its two sides---algebraic in the liquid and exponential in
the gas---makes this distinction sharper than the density profile alone.

\paragraph{Quasiparticle and hydrodynamic interpretation.}
The geometry has a simple quasiparticle interpretation. A spectral point
\(z\) labels a freely propagating quasiparticle in band \(\sigma\), whose
characteristic is
\begin{equation}
x_\sigma(t,z)=x_{0,\sigma}(z)+t\,v_\sigma(z).
\label{eq:discussion-free-flight}
\end{equation}
The velocity \(v_\sigma\) is fixed by the band dispersion, while the
effective midpoint \(x_{0,\sigma}(z)\) contains all information about the
return boundary condition and should not be identified with the bare
microscopic domain wall. In the exact solution it is encoded by the
normalized Abelian differential entering the Wiener--Hopf factors. The
arctic curves are the caustics of this family of trajectories,
\begin{equation}
\partial_zx_\sigma(t,z)=0.
\label{eq:discussion-caustic}
\end{equation}
This separation between quasiparticle motion and boundary data suggests
a formulation that may extend beyond free fermions, with the bare
velocities replaced by dressed ones. The noncompact real component of
the spectral curve produces the outer frozen--liquid boundary, whereas
its compact component produces the inner liquid--gas boundary. This
explains geometrically why the inner curve collapses together with the
compact component when the gap is closed. The hydrodynamic construction
of the midpoint curve and its relation to the Wiener--Hopf differentials
are detailed in Appendix~\ref{app:hydrodynamic-picture}.

\paragraph{Two domain walls and merging transitions.}
Local observables in the thermodynamic limit probe a single domain wall:
the second wall, at distance $N$, disappears after $N\to\infty$ at fixed
$R$. The return amplitude is instead a global quantity and retains both
ends of the occupied interval. Reflection combined with particle--hole symmetry makes the leading
free energies of the two walls equal, giving twice the single-wall
contribution to the return free energy. This separation also indicates a
natural extension of the present problem. At finite $N$, or in a double
scaling limit $N,R\to\infty$ with $N/R$ fixed, the two arctic regions no
longer decouple. Their collision gives the merger transition studied for
the homogeneous chain in Ref.~\cite{Pallister_2022}. In the two-band model
one may ask how the inner boundaries behave during this process: in
particular, whether two gas regions can merge, terminate, or be separated
by an intermediate liquid region. This should lead to a richer family of
merger transitions than in the one-band problem. Related collisions of
determinantal liquid regions lead to tacnode processes
\cite{adler2013nonintersecting,adler2014double}, suggesting a natural
universality question for the additional gas boundaries.

\paragraph{Relation to periodic dimers.}
The analytic structure of the solution closely parallels that of periodic
dimer models. There as well, matrix-valued weights lead to algebraic
spectral curves, and compact real components are associated with gas
phases. In the present continuous-time problem the matrix Wiener--Hopf
factorization reduces to a scalar Riemann--Hilbert problem on a genus-one
curve; the Abelian exponential, theta quotient, and moving divisor have
direct analogues in the two-periodic and doubly periodic Aztec-diamond
problems \cite{Boutillier2007PatternDensities,chhita2016domino,
duits2020two,berggren2019correlation,BerggrenBorodin2023}. The
factor-swapping construction of
Appendix~\ref{app:alternative-scalar-reduction-factor-swapping} makes the
relation especially transparent. Its continuous-time counterpart is the
isospectral factorization flow derived in
Appendix~\ref{app:factorization-flow}. We
expect this correspondence to extend to free-fermion chains with a larger
unit cell, where higher-genus curves and several compact components may
produce several incompressible regions and several inner arctic curves.

\paragraph{Extension to the Rice--Mele chain.}
The appearance of the gas phase is not specific to a gap opened by a
staggered on-site potential. As shown in
Appendix~\ref{app:rice-mele-extensions}, the dispersion of the full
Rice--Mele chain takes the same form as in the model studied in the main
text after introducing
\[
J=\sqrt{vw},
\qquad
\mu^2=\Delta^2+(v-w)^2.
\]
The elliptic spectral curve, leading return free energy, and Euler-scale
arctic geometry therefore carry over under the corresponding
reparametrization. In particular, the compact real component of the
spectral curve, and hence the incompressible gas region, is present
whenever \(\mu\neq0\), irrespective of whether the gap is opened by the
staggered potential, hopping dimerization, or both. The SSH chain is
recovered on the slice \(\Delta=0\). Although the leading scalar geometry
depends only on \(J\) and \(\mu\), sublattice-resolved observables can
retain additional dependence on the Bloch eigenvectors.

Several more immediate questions remain open. First, the analysis here is
a bulk analysis. A uniform treatment of the outer and inner arctic curves
requires local edge parametrices. A generic coalescence of two saddles is
described by the Chester--Friedman--Ursell reduction
\cite{Chester1957AnExtension} and suggests Airy scaling, as at the
ordinary arctic boundary \cite{Prahofer2002,Johansson2005}; the
liquid--gas edge may nevertheless retain additional band and matrix
structure. Second, the closing-gap regime is nonuniform:
the gas region shrinks while its correlation length diverges. The double
scaling limit $R\Delta^2=O(1)$ should describe the crossover between the
homogeneous arctic-circle geometry and a resolved gas region. Finally, the exact return amplitude \eqref{eq:return-amplitude-exact}
contains a bounded periodic theta-function correction to its leading
quadratic term. It would be interesting to determine how this
finite-gap structure generalizes to free-fermion chains with more
bands, and how it is related to the subleading emptiness and
full-counting-statistics asymptotics known for homogeneous
free-fermion chains
\cite{Franchini2005Asymptotics,Abanov2011QuantumFluctuations}. A complementary extension
is finite temperature, where even the homogeneous free Fermi gas exhibits
a transition between distinct mechanisms producing a large empty region
\cite{DelVecchio2026ATransition}. It would be interesting to understand how
this transition is modified by the band gap and the gas phase found here.

The broader conclusion is that the arctic geometry of a free-fermion
return problem reflects not only the ballistic motion of quasiparticles
but also the global geometry of their spectral curve. In the present
example the band gap, the compact spectral component, the incompressible
gas region, and the inner arctic curve are four descriptions of the same
structure. Matrix Wiener--Hopf factorization provides the link between
them.

\section*{Acknowledgements}

A.G.A. is grateful to Eldad Bettelheim for discussions of matrix
Riemann--Hilbert problems and to Alexei Borodin for discussions of
Wiener--Hopf factorization in the context of doubly periodic dimer models. 

The authors used various OpenAI models through ChatGPT and Codex for assistance
with manuscript editing, organization, and the examination of analytical
and numerical calculations. All outputs were checked by the authors.
Responsibility for the final text rests with the authors.

\paragraph{Funding information.}
D.M.G. acknowledges support from the Leverhulme Trust under Grant
No.~RPG-2024-401.

\addtocontents{toc}{\protect\setcounter{tocdepth}{1}}
\appendix

\section{Finite-rank sources and the transition kernel}
\label{app:finite-rank-sources}

Here we derive the generating identity \eqref{eq:finite-source-generating} and the finite-$N$ transition kernel \eqref{eq:finite-kernel-operator}. Let \(F\) be a finite-rank operator on the one-particle Hilbert space.
Its second quantization is
\[
\widehat F=
\sum_{m,n\in\mathbb Z}\sum_{a,b=1}^{2}
F_{ma,nb}\psi_{m,a}^\dagger\psi^{\phd}_{n,b},
\qquad
F_{ma,nb}=\langle m,a|F|n,b\rangle.
\]
Thus \(\widehat F\) is a number-conserving fermionic bilinear. Inserting $e^{\widehat F}$ at Euclidean time $\tau$ gives
\[
\mathcal Z_N[F;\tau]=
\langle\mathrm{DW}_N|
e^{-(R-\tau)\widehat H}e^{\widehat F}e^{-(R+\tau)\widehat H}
|\mathrm{DW}_N\rangle.
\]
On the one-particle space, $e^{\widehat F}$ acts as $e^F$. Since $|\mathrm{DW}_N\rangle$ is the Slater determinant associated with $\mathcal H_N$, the overlap is
\begin{equation}
\mathcal Z_N[F;\tau]=
\det_{\mathcal H_N}
\left[
\mathcal P_N\mathcal U_-e^F\mathcal U_+\mathcal P_N
\right],
\qquad
\mathcal U_-=\mathcal U(R-\tau),
\quad
\mathcal U_+=\mathcal U(R+\tau).
\label{eq:app-source-overlap}
\end{equation}
The source-free overlap matrix is
\[
A_N=
\mathcal P_N\mathcal U_-\mathcal U_+\mathcal P_N\big|_{\mathcal H_N}
=T_N[\mathbf W_R].
\]
Writing $\mathcal Q=e^F-I$, the matrix in \eqref{eq:app-source-overlap} becomes
\[
A_N[F]=A_N+\mathcal P_N\mathcal U_-\mathcal Q\mathcal U_+\mathcal P_N.
\]
Factoring out $A_N$ therefore gives
\[
\frac{\mathcal Z_N[F;\tau]}{Z_N(R)}
=
\det_{\mathcal H_N}
\left[
I+A_N^{-1}\mathcal P^{\phd}_N\mathcal U^{\phd}_-\mathcal Q\mathcal U^{\phd}_+\mathcal P^{\phd}_N
\right].
\]
Because $\mathcal Q$ has finite rank, Sylvester's identity
$\det(I+XY)=\det(I+YX)$ transfers this determinant to the
finite-dimensional support of $\mathcal Q$. Hence
\[
\frac{\mathcal Z_N[F;\tau]}{Z_N(R)}
=
\det\left[I+\mathcal Q\mathcal C_N(\tau)\right],
\qquad
\mathcal C_N(\tau)=
\mathcal U^{\phd}_+\mathcal P^{\phd}_NA_N^{-1}\mathcal P^{\phd}_N\mathcal U^{\phd}_-,
\]
which is \eqref{eq:finite-source-generating}.

To extract the transition kernel, choose
\[
F=\lambda|n,b\rangle\langle m,a|.
\]
Then
\[
\mathcal Q=e^F-I
=
\lambda|n,b\rangle\langle m,a|+O(\lambda^2),
\]
and differentiation at $\lambda=0$ gives
\[
\left.
\partial_\lambda\log\mathcal Z_N[F;\tau]
\right|_{\lambda=0}
=
\langle m,a|\mathcal C_N(\tau)|n,b\rangle.
\]
The left-hand side is the normalized matrix element defining
$[\mathbf K_N(m,n;\tau)]_{ab}$, proving
\eqref{eq:finite-kernel-operator}. In components,
\begin{align}
[\mathbf K_N(m,n;\tau)]_{ab}
={}&
\sum_{\ell,j=0}^{N-1}
\sum_{c,d=1}^{2}
\langle m,a|\mathcal U(R+\tau)|\ell,c\rangle
\left(T_N[\mathbf W_R]^{-1}\right)_{\ell c,jd}
\langle j,d|\mathcal U(R-\tau)|n,b\rangle.
\label{eq:app-finite-kernel-components}
\end{align}
Higher derivatives with respect to finite-rank sources generate multipoint fermion correlators. Expanding the determinant reproduces the usual determinantal Wick formulas, so all normalized observables are encoded in $\mathcal C_N(\tau)$.

\section{Toeplitz consequences of the canonical factors}
\label{app:WH-Toeplitz}

This appendix supplies the operator details used at the beginning of
Section~\ref{sec:WienerHopf}. They are independent of the explicit
elliptic-curve construction. To simplify the formulas, we suppress the
spectral argument \(z\) and the parameter \(R\) in
\(\boldsymbol\Gamma_\pm(z;R)\) whenever they are clear from context.

\subsection{The inverse Toeplitz operator}

Consider the dual canonical factorization
\[
\mathbf W_R=\boldsymbol\Gamma_+\boldsymbol\Gamma_-^{-1}.
\]
The analyticity of $\boldsymbol\Gamma_+^{\pm1}$ inside the unit circle and of
$\boldsymbol\Gamma_-^{\pm1}$ outside it implies the exact half-line Toeplitz
product identities
\[
\mathsf T[\mathbf W_R]=\mathsf T[\boldsymbol\Gamma_+]\mathsf T[\boldsymbol\Gamma_-^{-1}],
\qquad
\mathsf T[\mathbf W_R]^{-1}=\mathsf T[\boldsymbol\Gamma_-]\mathsf T[\boldsymbol\Gamma_+^{-1}].
\]
Writing the second identity in Fourier components gives
\[
\bigl(\mathsf T[\mathbf W_R]^{-1}\bigr)_{\ell b,ja}
=
\sum_{r\geq0}\sum_{c=1}^{2}
(\boldsymbol\Gamma_-)_{\ell-r,bc}
(\boldsymbol\Gamma_+^{-1})_{r-j,ca}.
\]
Multiplication by $w^{-\ell}z^j$ and summation over
$\ell,j\geq0$ produces the geometric series
\[
\sum_{r\geq0}\left(\frac zw\right)^r=\frac{w}{w-z},
\qquad |z|<|w|,
\]
and hence
\[
\sum_{\ell,j\geq0}w^{-\ell}z^j
\bigl(\mathsf T[\mathbf W_R]^{-1}\bigr)_{\ell b,ja}
=
\left[
\frac{w}{w-z}\boldsymbol\Gamma_-(w)\boldsymbol\Gamma_+(z)^{-1}
\right]_{ba}.
\]
This proves Eq.~\eqref{eq:WH-inverse-generating}. The Cauchy factor is
universal: it is the generating function of the identity on the
half-line. All model dependence is contained in the two
Wiener--Hopf factors.

\subsection{The transition kernel}

In the thermodynamic limit defining \(\mathbf K(m,n;\tau)\) in Eq.~\eqref{eq:thermodynamic-kernel}, the one-particle operator underlying the kernel becomes
\[
\mathcal C(\tau)
=
\mathcal U(R+\tau)\mathcal P_+
\mathsf T[\mathbf W_R]^{-1}\mathcal P_+
\mathcal U(R-\tau),
\]
where \(\mathcal P_+\) projects onto nonnegative cell indices. Substituting the
generating function above and Fourier transforming the two external
propagators gives
\begin{align}
\mathbf K(m,n;\tau)
={}&
\oint_{|z|<|w|}
\frac{dw}{2\pi iw}\frac{dz}{2\pi iz}\,
w^m z^{-n}\frac{w}{w-z}
\nonumber\\[-1mm]
&\hspace{9mm}\times
\mathbf U(R+\tau;w)\boldsymbol\Gamma_-(w)
\boldsymbol\Gamma_+(z)^{-1}\mathbf U(R-\tau;z).
\label{eq:app-K-before-jump}
\end{align}
From $\mathbf W_R=\boldsymbol\Gamma_+\boldsymbol\Gamma_-^{-1}$ one has
\[
\boldsymbol\Gamma_+^{-1}
=\boldsymbol\Gamma_-^{-1}\mathbf W_R^{-1}
=\boldsymbol\Gamma_-^{-1}\mathbf U(-2R;z).
\]
Substituting this identity into
Eq.~\eqref{eq:app-K-before-jump} gives the exact kernel
representation \eqref{eq:K1} used in Section~\ref{sec:saddle-geometry}.
This is a matrix identity in the sublattice space. Taking $n=m$ and a
diagonal entry gives the exact thermodynamic-limit density.

\subsection{The Widom differential}

We now derive the specialized differential identity stated in
Eq.~\eqref{eq:results-Widom-differential}. At finite \(N\), Jacobi's
identity gives
\[
\partial_R\log Z_N(R)
=\operatorname{Tr}\left[
T_N[\mathbf W_R]^{-1}T_N[\partial_R\mathbf W_R]
\right].
\]
The same inverse Toeplitz matrix therefore appears in both observables:
a finite-rank insertion selects its matrix elements, whereas changing
$R$ tests it against an extended Toeplitz perturbation. Since
$\det \mathbf W_R=1$, the extensive Szeg\H{o} contribution vanishes. The regularized
$N\to\infty$ limit of Jacobi's identity is Widom's differential formula,
\[
\partial_R\log Z(R)
=\frac{1}{2\pi i}\oint_{|z|=1}
\operatorname{Tr}\Bigl[
\mathbf W_R^{-1}\partial_R\mathbf W_R
\bigl(
\partial_z\boldsymbol\Gamma_-\boldsymbol\Gamma_-^{-1}
+\boldsymbol\Psi_+^{-1}\partial_z\boldsymbol\Psi_+
\bigr)
\Bigr]\,dz,
\]
The exterior term comes from the dual factorization and the interior term
from the direct one; they are the two edge contributions to the
regularized determinant.

For the present symbol the direct factors need not be constructed
independently. Since $\mathbf W_R^{-1}=\mathbf W_{-R}$, uniqueness of the normalized
factorization gives
\begin{equation}
\boldsymbol\Psi_+(z;R)=\boldsymbol\Gamma_+(0;-R)\boldsymbol\Gamma_+(z;-R)^{-1}.
\label{eq:app-direct-from-dual}
\end{equation}
The constant prefactor drops out after differentiation in $z$. Moreover,
$\boldsymbol\Gamma_+(z;s)=\mathbf W_s(z)\boldsymbol\Gamma_-(z;s)$ and therefore
\begin{align}
\frac{1}{2\pi i}\oint
\operatorname{Tr}\left[\mathbf h\,\boldsymbol\Psi_+^{-1}d\boldsymbol\Psi_+\right]
=-\frac{1}{2\pi i}\oint
\operatorname{Tr}\left[\mathbf h\,d\boldsymbol\Gamma_-(-R)\boldsymbol\Gamma_-(-R)^{-1}\right].
\label{eq:app-reflected-edge}
\end{align}
Here the additional term generated by differentiating $\mathbf W_{-R}$ integrates
to zero, because
\[
\operatorname{Tr}\left[\mathbf h\,d\mathbf W_s\mathbf W_s^{-1}\right]
=-s\,d\operatorname{Tr}\mathbf h^2.
\]
Finally, using $\mathbf W_R^{-1}\partial_R\mathbf W_R=-2\mathbf h$ in the standard Widom formula
gives Eq.~\eqref{eq:WH-Widom-differential}. Thus both edge contributions
are expressed through $\boldsymbol\Gamma_-$, while only the factors at $R$ are needed
for the inverse Toeplitz operator.

\section{Analytic details of the elliptic Wiener--Hopf factors}
\label{app:WH-analytic-details}

Here we verify the analytic properties of the construction in
Section~\ref{sec:WienerHopf}. In particular, we explain the split of the
essential singularities, the cancellation of monodromy, and the removal
of the auxiliary divisor.

\subsection{Normalized differentials and the singularities of the energy}

Near the branch points $P_0$ and $P_\infty$, take local coordinates
$u_0=z^{1/2}$ and $u_\infty=z^{-1/2}$, respectively. The energy has the
expansions
\[
E(P)=\frac{1}{u_0}+\frac{A}{2}u_0+O(u_0^3),
\qquad
E(P)=\frac{1}{u_\infty}+\frac{A}{2}u_\infty+O(u_\infty^3).
\]
Thus a naive assignment $\chi_+=e^{-2RE}$, $\chi_-=1$ would put an
essential singularity of the interior factor at $P_0$. Instead, define
the second-kind differentials in Eq.~\eqref{eq:second-kind-WH}. A direct
calculation using $E=\eta/z$ and $\eta^2=f(z)$ gives
\[
d\Omega_\infty+d\Omega_0=-2\,dE.
\]
The common constant $c$ is fixed by the vanishing $a$-period:
\[
0=\oint_a d\Omega_\infty
=-\oint_a\frac{z-c}{\eta}\,dz,
\]
which gives Eq.~\eqref{eq:c-WH}; the $a$-period of $d\Omega_0$ then
vanishes as well. Additive constants in the Abelian integrals may be
chosen so that
\[
\Omega_\infty(P)+\Omega_0(P)=-2E(P).
\]
The function $\Omega_\infty$ is regular at $P_0$ and has the required
principal part at $P_\infty$, while $\Omega_0$ is regular at $P_\infty$
and has the required principal part at $P_0$. It follows immediately
that the two exponentials in Eq.~\eqref{eq:scalar-WH-solution} have the
jump \eqref{eq:scalar-WH-jump} and the correct local analyticity.

\subsection{Monodromy and the theta quotient}

The zero $a$-period makes the Abelian exponentials single-valued around
the $a$-cycle. Their $b$-cycle monodromy is determined by
\[
V=\frac{1}{2\pi i}\oint_b d\Omega_\infty.
\]
The genus-one theta function obeys
\[
\theta(u+1|B)=\theta(u|B),
\qquad
\theta(u+B|B)=e^{-\pi iB-2\pi iu}\theta(u|B).
\]
Together with the corresponding period relation for the normalized
second-kind differential, these identities show that the theta quotient
\eqref{eq:theta-quotient-WH} has precisely the inverse multiplier of
the Abelian exponential around the $b$-cycle. The products
$\widetilde\chi_\pm$ are therefore single-valued.

The theta function has one zero at
\[
K_B=\frac{1+B}{2}
\pmod{\mathbb Z+B\mathbb Z}.
\]
Consequently, $q(P;R)$ has a fixed pole at $P_D$ and a moving zero
$P_Z(R)$ satisfying
\[
u(P_Z(R))\equiv u(P_D)-RV
\pmod{\mathbb Z+B\mathbb Z}.
\]
This divisor is auxiliary: it is the price of cancelling the monodromy
and must disappear from the final matrix factors.

\subsection{Removal of the auxiliary divisor}

Reconstruct the preliminary matrix factors as
\[
\widetilde{\boldsymbol\Gamma}_\pm
=\widetilde\chi_\pm(P)\mathbf Y(P)
 +\widetilde\chi_\pm(P^*)\mathbf Y(P^*).
\]
At $z_D=\pi(P_D)$ they have a rank-one pole with left row direction
$\ell_D^\top$, while at $z_Z=\pi(P_Z)$ their determinant has a simple
zero with right kernel spanned by $v_Z$. Assume that neither point is a
branch point and that $\ell_D^\top v_Z\neq0$. Introduce
\[
\boldsymbol\Pi=\frac{v_Z\ell_D^\top}{\ell_D^\top v_Z},
\qquad
\mathbf C(z;R)=I+\frac{z_Z-z_D}{z-z_Z}\boldsymbol\Pi.
\]
The rank-one matrix $\boldsymbol\Pi$ satisfies
\[
\boldsymbol\Pi^2=\boldsymbol\Pi,
\qquad
\boldsymbol\Pi v_Z=v_Z,
\qquad
\ell_D^\top(I-\boldsymbol\Pi)=0.
\]
Near $z_D$, the last identity removes the row direction of the residue
in $\widetilde{\boldsymbol\Gamma}_\pm\mathbf C$. Near $z_Z$, the apparent pole of
$\mathbf C$ acts
only in the direction $v_Z$, which belongs to the right kernel of
$\widetilde{\boldsymbol\Gamma}_\pm(z_Z)$; its residue therefore vanishes. Hence
$\boldsymbol\Gamma_\pm^{\mathrm{corr}}=\widetilde{\boldsymbol\Gamma}_\pm\mathbf C$ has neither the pole at
$z_D$ nor the determinant zero at $z_Z$. Since $\mathbf C$ is the same on both
sides of the contour, it does not change the jump.

The constant right normalization
\[
\boldsymbol\Gamma_\pm(z;R)
=\boldsymbol\Gamma_\pm^{\mathrm{corr}}(z;R)
\boldsymbol\Gamma_-^{\mathrm{corr}}(\infty;R)^{-1}
\]
then gives $\boldsymbol\Gamma_-(\infty;R)=I$ and preserves
$\mathbf W_R=\boldsymbol\Gamma_+\boldsymbol\Gamma_-^{-1}$. This proves
Eq.~\eqref{eq:exact-dual-WH-factors}. Finally,
$\mathbf W_R^{-1}=\mathbf W_{-R}$ implies
\[
\mathbf W_R
=\boldsymbol\Gamma_-(z;-R)\boldsymbol\Gamma_+(z;-R)^{-1},
\]
and the normalization at the origin gives
Eq.~\eqref{eq:app-direct-from-dual}.

\section{Direct evaluation of the return amplitude}
\label{app:return-finite-gap}

We evaluate the return amplitude exactly and show directly that its
logarithm has no term linear in \(R\). We assume \(\Delta\neq0\) and use
the notation of Section~\ref{sec:WienerHopf}, with real cycles chosen so
that \(B=i\beta\), \(\beta>0\), and \(V\in\mathbb R\).

\subsection{Extracting one coefficient of the matrix factor}

Write the normalized exterior factor as
\[
\boldsymbol\Gamma_-(z;R)
=I+\frac{\boldsymbol\Gamma_1(R)}{z}+O(z^{-2}),
\qquad
g(R)=[\boldsymbol\Gamma_1(R)]_{12}.
\]
Since the only growing entry of \(\mathbf h(z)\) is
\(\mathbf h_{21}(z)=z+1\), evaluating the contour integral in
Eq.~\eqref{eq:return-Widom-short} at infinity gives
\begin{equation}
\partial_R\log Z(R)=2\bigl[g(R)-g(-R)\bigr].
\label{eq:return-short-Widom}
\end{equation}

By uniqueness of the canonical factorization, the normalized matrix
factors are independent of the auxiliary point \(P_D\). We choose the
limit \(P_D\to P_*=(-1,-\Delta)\), approaching from admissible points
away from the jump contour. At this point
\(\mathbf Y(P_*)=\operatorname{diag}(1,0)\), so the row direction of the
auxiliary pole is \(\ell_D^\top=(1,0)\). The rational correction
\eqref{eq:rational-correction-WH} therefore satisfies
\(\mathbf C(z;R)e_2=e_2\), where \(e_2=(0,1)^\top\): it leaves the
second column unchanged.

Near \(P_\infty\), set \(s=z^{-1/2}\) and expand the scalar factor as
\[
\widetilde\chi_-(P;R)
=\chi_\infty(R)\bigl[1+a(R)s+O(s^2)\bigr].
\]
Sheet exchange sends \(s\mapsto-s\), and
\((\mathbf h/E)_{12}=s+O(s^3)\). The reconstruction formula
\eqref{eq:preliminary-matrix-WH} thus gives
\([
\widetilde{\boldsymbol\Gamma}_-(z;R)]_{12}
=\chi_\infty(R)a(R)/z+O(z^{-2})\) and
\(\widetilde{\boldsymbol\Gamma}_-(\infty;R)e_2
=\chi_\infty(R)e_2\). The normalization
\eqref{eq:exact-dual-WH-factors} then yields \(g(R)=a(R)\).

The local expansions of the Abelian integrals are
\[
u(P)=u(P_\infty)-\frac{2}{\mathcal A}s+O(s^3),
\qquad
\Omega_0(P)=\Omega_0(P_\infty)+2cs+O(s^3).
\]
The Riemann bilinear relation gives
\(V=\operatorname*{Res}_{P_\infty}(u\,d\Omega_\infty)
=-4/\mathcal A\). Substituting these expressions into the scalar
solution \eqref{eq:scalar-WH-solution}, we obtain
\begin{equation}
g(R)=-2cR+\frac{V}{2}
\left[
\frac{\theta'(D+RV\,|\,B)}{\theta(D+RV\,|\,B)}
-\frac{\theta'(D\,|\,B)}{\theta(D\,|\,B)}
\right],
\label{eq:return-short-g}
\end{equation}
where \(D=u(P_\infty)+K_B-u(P_*)\), and the prime denotes the
derivative with respect to the first argument.

\subsection{The theta correction and its boundedness}

Inserting Eq.~\eqref{eq:return-short-g} into
Eq.~\eqref{eq:return-short-Widom} cancels the \(R\)-independent term.
The remaining theta contribution is
\(\partial_R\log[\theta(D+RV\,|\,B)\theta(D-RV\,|\,B)]\).
Integration with \(Z(0)=1\) gives
\begin{equation}
Z(R)=e^{-4cR^2}
\frac{\theta(D+RV\,|\,B)\theta(D-RV\,|\,B)}
     {\theta(D\,|\,B)^2}.
\label{eq:return-short-theta-product}
\end{equation}

The shift \(D\) is fixed by the position of \(P_*\). The inversion
\(\iota:(z,\eta)\mapsto(z^{-1},\eta/z^2)\) exchanges \(P_0\) and
\(P_\infty\), fixes \(P_*\), and reverses \(d\omega\). With
\(w(P)=u(P)-u(P_\infty)\), the real path from \(P_\infty\) to
\(P_0\) is half an \(a\)-cycle, giving \(w(P_0)\equiv1/2\). Hence
\(w(\iota P)\equiv1/2-w(P)\) and \(2w(P_*)\equiv1/2\).
Since \(P_*\) lies on the compact real component, where
\(\operatorname{Im}w=\beta/2\), it follows that
\[
w(P_*)\equiv\frac{B}{2}\pm\frac14,
\qquad D\equiv\pm\frac14
\pmod{\mathbb Z+B\mathbb Z}.
\]
The normalized product in Eq.~\eqref{eq:return-short-theta-product}
is invariant under period shifts of \(D\). Taking either representative
and applying the theta duplication identity,
\[
\frac{\theta(\tfrac14+y\,|\,B)\theta(\tfrac14-y\,|\,B)}
     {\theta(\tfrac14\,|\,B)^2}
=\frac{\theta(2y\,|\,2B)}{\theta(0\,|\,2B)},
\]
we obtain the exact return amplitude
\begin{equation}
\boxed{
Z(R)=e^{-4cR^2}
\frac{\theta(2RV\,|\,2B)}{\theta(0\,|\,2B)}.
}
\label{eq:return-short-exact}
\end{equation}

For \(B=i\beta\), \(\beta>0\), the Jacobi product formula implies
\(\theta(y\,|\,2B)>0\) for real \(y\). This function is continuous
and periodic, so it has a strictly positive minimum and a finite
maximum on the real axis. The logarithm of the theta ratio in
Eq.~\eqref{eq:return-short-exact} is therefore bounded and periodic in
\(R\), with period \(1/(2|V|)\). Consequently, at fixed
\(\Delta\neq0\),
\[
\log Z(R)=-4cR^2+O(1),
\]
which establishes the absence of a linear term directly.

For numerical evaluation, the periods can be expressed directly in
terms of complete elliptic integrals. Set
\[
m_{\mathrm{ell}}=1-\kappa^{-2}.
\]
With the cycle orientations used above,
\begin{equation}
\mathcal A
=\frac{4}{\sqrt{\kappa}}\,\mathrm K(m_{\mathrm{ell}}),
\qquad
B
=i\,\frac{\mathrm K(1-m_{\mathrm{ell}})}
          {\mathrm K(m_{\mathrm{ell}})},
\qquad
V
=-\frac{4}{\mathcal A}
=-\frac{\sqrt{\kappa}}{\mathrm K(m_{\mathrm{ell}})}.
\label{eq:explicit-elliptic-periods}
\end{equation}
Here \(\mathrm K\) is the complete elliptic integral of the first kind.
The bounded correction in Eq.~\eqref{eq:return-short-exact} therefore
has period
\[
T_R=\frac{1}{2|V|}
=\frac{\mathrm K(1-\kappa^{-2})}{2\sqrt{\kappa}}.
\]
Using these expressions for \(B\) and \(V\), we plot the finite-gap
correction for three representative values of \(\Delta\) in
Fig.~\ref{fig:return-amplitude-remainder}.
The sign of \(V\) depends on the cycle orientation and does not affect
the theta quotient. In the closing-gap limit.  In the closing-gap limit
\(\Delta\to0\), one has \(B\to i\infty\) and
\(V\to-2/\pi\); the theta quotient tends to unity, so the bounded
correction vanishes.

\section{Steepest-descent contours and phase topology}
\label{app:steepest-descent-topology}

This appendix gives the contour details behind
Sections~\ref{sec:saddle-geometry} and
\ref{sec:transition-kernel}. We first recover the spectral-curve
integral from the exact kernel, then describe how the Cauchy-pole residue
changes across the phase diagram.

\subsection{From band indices to the spectral curve}

Resolving the Wiener--Hopf factors in Eq.~\eqref{eq:K1} into their two
spectral components, the scaled finite-separation kernel takes the form
\begin{align}
\mathbf K(M+m_0,M+n_0;tR)
={}&
\sum_{\sigma,\lambda=\pm}
\int_{\Gamma_{w,\sigma}}\frac{dw}{2\pi iw}
\int_{\Gamma_{z,\lambda}}\frac{dz}{2\pi iz}\,
\frac{w}{w-z}\,w^{m_0}z^{-n_0}
\nonumber\\
&\quad\times
e^{R[S_\sigma(w;x_R,t)-S_\lambda(z;x_R,t)]}
\mathbf A_{\sigma\lambda}(w,z),
\label{eq:app-sheet-kernel}
\end{align}
where
\begin{align}
S_\sigma(z;x,t)
&=x\log z-\Omega_{0,\sigma}(z)-(1+t)E_\sigma(z),\\
\mathbf A_{\sigma\lambda}(w,z)
&=\frac{q_\sigma(w;R)}{q_\lambda(z;R)}
\mathbf Y_\sigma(w)\mathbf C(w;R)\mathbf C(z;R)^{-1}\mathbf Y_\lambda(z).
\label{eq:app-kernel-amplitude}
\end{align}
Here the subscript \(\sigma\) denotes evaluation at the point
\(P_\sigma=(z,\sigma\eta(z))\) on sheet \(\sigma\), and similarly for
the \(w\)-dependent quantities. The $w$-contours enclose the corresponding $z$-contours. The union of
the two band contours is precisely a contour on the two-sheeted curve:
\[
\Gamma_P=\Gamma_{w,+}\cup\Gamma_{w,-},
\qquad
\Gamma_Q=\Gamma_{z,+}\cup\Gamma_{z,-}.
\]
Replacing the sheet label by the point $P$ or $Q$ gives
Eq.~\eqref{eq:spectral-double-kernel}. All theta and rational factors are retained in
\(\mathbf A_{\sigma\lambda}\), but they do not modify the
large-\(R\) saddle action.

\subsection{Opposite steepest-descent deformations}

Set
\[
H(P;x,t)=\operatorname{Re}S(P;x,t).
\]
The magnitude of the exponential in
Eq.~\eqref{eq:spectral-double-kernel} is
\[
e^{R[H(P)-H(Q)]}.
\]
The $P$-contour is therefore deformed along decreasing directions of
$H$, while the $Q$-contour is deformed along increasing directions.
Near a simple saddle $P_c$, the holomorphic Morse lemma gives a local
coordinate $\zeta$ such that
\[
S(P)-S(P_c)=\frac{\zeta(P)^2}{2}.
\]
The two descent and two ascent rays alternate at $P_c$. The global
contours are assembled from these rays and from trajectories ending at
$P_0$ or $P_\infty$.

For the present genus-one phase portrait, the ascent and descent graphs can be followed continuously from the real ovals throughout each connected bulk region. The resulting contours satisfy \(H(P)<H(Q)\) away from isolated common saddles. Their homology and relative nesting cannot change within a bulk region: a change requires the coalescence of critical points, which occurs precisely on the double-root locus defining the arctic curves. This gives the required steepest-descent deformation in each of the five bulk regions. A uniform analysis at a double root would require a separate edge parametrix and is outside the scope of the present bulk calculation.

Before estimating the remaining double integral, any crossing of the Cauchy pole \(w=z\) is extracted as a residue. After this subtraction the two contours may meet only at isolated simple saddles, where the remaining prefactor is regular. The two Gaussian integrations then give an \(O(R^{-1})\) contribution. An order-one term arises only from the extracted Cauchy-pole residue,
\[
\int_{\gamma(x,t)}
\frac{dz(P)}{2\pi iz(P)}\,z(P)^r\mathbf Y(P),
\]
because at coincident points the corrected prefactor reduces to the spectral projector. This proves
Eq.~\eqref{eq:local-kernel-contour}.

\subsection{Homology of the swept contour}

The contour $\gamma(x,t)$ can be determined by continuity across the
phase diagram.

\begin{itemize}
\item Far in the left frozen region the original nesting is unchanged,
so $\gamma$ is empty and $\mathbf K=0$.

\item At the left frozen--liquid curve, two positive-real saddles
coalesce and become a complex-conjugate pair. The swept contour is the
lower-sheet arc joining this pair. Writing the endpoints as
$\varrho e^{\pm ik_F}$ gives
Eq.~\eqref{eq:left-liquid-kernel}.

\item At the liquid--gas curve the endpoints meet on the compact
negative oval. The occupied arc closes into the full lower-band cycle,
giving Eq.~\eqref{eq:gas-kernel-contour}.

\item Across the right liquid--gas curve it is simpler to use the
complementary description. A hole arc opens on the upper sheet, giving
Eq.~\eqref{eq:right-liquid-kernel}.

\item At the right frozen--liquid curve the hole arc collapses. Both
bands are then filled and $\mathbf K(r)=\delta_{r0}I$.
\end{itemize}

These five homology classes give the piecewise kernel
\eqref{eq:piecewise-local-kernel}. They also explain why the gas state
is position independent: once the contour is a complete band cycle,
its homology no longer changes inside the gas region.

\section{Evaluation of local kernels and correlation asymptotics}
\label{app:kernel-correlation-evaluation}

Here we evaluate the contour formulas of
Section~\ref{sec:transition-kernel} in components and derive their
large-distance behavior.

\subsection{Component form of the liquid and gas kernels}

For either the occupied or hole arc, let
\begin{align}
s_r(k_F,\varrho)
&=\int_{-k_F}^{k_F}\frac{dk}{2\pi}\,
\varrho^r e^{irk}
=
\begin{cases}
k_F/\pi,&r=0,\\
\varrho^r\sin(rk_F)/(\pi r),&r\neq0,
\end{cases}\\
b_r^{(\sigma)}
&=\varrho^r\int_{-k_F}^{k_F}\frac{dk}{2\pi}\,
\frac{e^{ikr}}{E_\sigma(k)},
\qquad
\sigma=-,+.
\end{align}
The arc integral of the projector is
\begin{equation}
\mathbf L_\sigma(r;x,t)
=\frac12
\begin{pmatrix}
s_r+\Delta b_r^{(\sigma)}
&
b_{r-1}^{(\sigma)}+b_r^{(\sigma)}
\\
b_r^{(\sigma)}+b_{r+1}^{(\sigma)}
&
s_r-\Delta b_r^{(\sigma)}
\end{pmatrix}.
\label{eq:app-liquid-component-kernel}
\end{equation}
Thus the left-liquid kernel is $\mathbf L_-$, while the
right-liquid kernel is
$\delta_{r0}I-\mathbf L_+$.

For the gas phase define
\[
\varepsilon(k)=\sqrt{\Delta^2+2+2\cos k},
\qquad
a_r=\int_{-\pi}^{\pi}\frac{dk}{2\pi}\,
\frac{e^{ikr}}{\varepsilon(k)}.
\]
Equation~\eqref{eq:gas-kernel-contour} becomes
\begin{equation}
\mathbf G(r)=\frac12
\begin{pmatrix}
\delta_{r0}-\Delta a_r&-a_{r-1}-a_r\\
-a_r-a_{r+1}&\delta_{r0}+\Delta a_r
\end{pmatrix}.
\label{eq:app-gas-component-kernel}
\end{equation}
At $r=0$, Eqs.~\eqref{eq:app-liquid-component-kernel} and
\eqref{eq:app-gas-component-kernel} give
Eq.~\eqref{eq:piecewise-density}. In particular,
\[
\rho_1^{\mathrm{gas}}
=\frac12
-\frac{\Delta}{\pi\sqrt{\Delta^2+4}}\,
\mathrm K\!\left(\frac{4}{\Delta^2+4}\right),
\qquad
\rho_2^{\mathrm{gas}}
=\frac12
+\frac{\Delta}{\pi\sqrt{\Delta^2+4}}\,
\mathrm K\!\left(\frac{4}{\Delta^2+4}\right).
\]

\subsection{Endpoint asymptotics in the liquid}

For a smooth liquid arc, integration by parts gives
\begin{equation}
b_r^{(\sigma)}
=\frac{\varrho^r}{2\pi ir}
\left[
\frac{e^{irk_F}}{E_{\sigma,F}}
-\frac{e^{-irk_F}}{\overline{E_{\sigma,F}}}
\right]
+O\!\left(\frac{\varrho^r}{r^2}\right),
\label{eq:app-br-endpoint}
\end{equation}
where $E_{\sigma,F}=E_\sigma(k_F)$ and $P_{\sigma,F}=P_\sigma(k_F)$.
Equivalently,
\begin{equation}
\mathbf L_\sigma(r;x,t)
=\frac{\varrho^r}{2\pi ir}
\left[
e^{irk_F}\mathbf Y(P_{\sigma,F})
-e^{-irk_F}\mathbf Y(\overline{P_{\sigma,F}})
\right]
+O\!\left(\frac{\varrho^r}{r^2}\right).
\label{eq:app-liquid-endpoint-kernel}
\end{equation}
The factors $\varrho^{\pm r}$ cancel between $\mathbf K(r)$ and $\mathbf K(-r)$.
Substitution in Eq.~\eqref{eq:cell-correlation-kernel} yields
Eq.~\eqref{eq:liquid-correlation-section8}, with
\[
T_F=\operatorname{Tr}\!\left[
\mathbf Y(P_F)\mathbf Y(\overline{P_F})
\right].
\]
Writing $z_F=\varrho e^{ik_F}$ and $E_F=\eta_F/z_F$, one finds
\begin{equation}
T_F
=\frac12+
\frac{
\Delta^2+1+(\varrho+\varrho^{-1})\cos k_F+\cos(2k_F)}
{2|E_F|^2}.
\label{eq:app-TF-explicit}
\end{equation}
At the time-symmetric slice $t=0$, $\varrho=1$ and this reduces to
\[
C_{\mathrm{cell}}^{\mathrm{liquid}}(r;x,0)
=-\frac{1}{2\pi^2r^2}
\left[
1-\left(1-\frac{\sin^2k_F}{\varepsilon_F^2}\right)
\cos(2rk_F)
\right]
+O(r^{-3}),
\]
where $\varepsilon_F=\varepsilon(k_F)$.

\subsection{Branch-point asymptotics in the gas}

The Fourier coefficients $a_r$ may be written as contour integrals in
$z=e^{ik}$. Their asymptotics follow from standard singularity analysis
\cite{Flajolet1990SingularityAnalysis}. Their nearest singularities are the square-root branch
points $z=-\kappa$ and $z=-\kappa^{-1}$. For $r>0$, deforming the
contour toward the inner branch point $-\kappa^{-1}$ gives the standard
Darboux behavior
\[
a_r=(-1)^r\,\kappa^{-r}\,O(r^{-1/2}),
\]
with the analogous expression for $r<0$. Substituting the local
square-root expansion into all four entries of
Eq.~\eqref{eq:app-gas-component-kernel} and then forming
$-\operatorname{Tr}[\mathbf G(r)\mathbf G(-r)]$ fixes both the power and the prefactor:
\[
C_{\mathrm{cell}}^{\mathrm{gas}}(r)
=-\frac{\kappa-\kappa^{-1}}{4\pi|r|}
\kappa^{-2|r|}
\left[1+O(|r|^{-1})\right].
\]
This is Eq.~\eqref{eq:gas-correlation-section8}. The exponential factor
gives
\[
\xi_{\mathrm{gas}}^{-1}=2\log\kappa
=4\operatorname{arsinh}(|\Delta|/2).
\]
The correlation length diverges as $\Delta\to0$, while the gas region
simultaneously collapses. A parametrically wide bulk-gas regime therefore requires
\[
R\Delta^2\gg1.
\]
When \(R\Delta^2=O(1)\), the gas width and its correlation length are
comparable.

\section{Factorization-flow derivation of the Wiener--Hopf factors}
\label{app:factorization-flow}

The main text constructs the Wiener--Hopf factors by solving a scalar
Riemann--Hilbert problem on the spectral curve. Here we give an alternative
derivation based on the evolution of the factors with the parameter $R$.
The method is the factorization-flow, or dressing, formulation of the
Adler--Kostant--Symes scheme
\cite{Adler1979,Kostant1979,SemenovTianShansky1983}. It does not solve the
matrix Riemann--Hilbert problem directly: once a canonical factorization
exists, differentiation turns it into an initial-value problem.

We work throughout with the factors used in the main text,
\begin{equation}
\mathbf W_R(z)=e^{-2R\mathbf h(z)}=\boldsymbol\Gamma_+(z;R)\boldsymbol\Gamma_-(z;R)^{-1},
\qquad \boldsymbol\Gamma_-(\infty;R)=I.
\label{eq:flow-WH}
\end{equation}
The scalarization below concerns the action of the factors on an
eigenline. It does not assume that $\boldsymbol\Gamma_\pm$ commute with the fixed
spectral projectors of $\mathbf h(z)$; in fact, the eigenlines move under the
flow.

\subsection{Canonical factorization as an initial-value problem}

Let $\mathsf P_+$ and $\mathsf P_-$ project a matrix Laurent series onto its nonnegative
and strictly negative powers of $z$, respectively. Their ranges are the
matrix functions holomorphic inside the unit circle and those holomorphic
outside and vanishing at infinity.

For real $s$, the symbol $e^{s\mathbf h(e^{ik})}$ is Hermitian positive-definite.
Its partial indices therefore vanish, and it has a canonical
factorization.\footnote{For a general matrix symbol a diagonal factor
$\operatorname{diag}(z^{\nu_1},\ldots,z^{\nu_m})$ may be required. The
integers $\nu_j$ are the partial indices; their vanishing is precisely the
condition for a canonical factorization.} Canonical factors depend
real-analytically on $s$ as long as the partial indices remain fixed
\cite{GohbergKrein1958,ClanceyGohberg1981}. We may therefore write
\begin{equation}
e^{s\mathbf h(z)}=\mathbf G_-(s,z)\mathbf G_+(s,z),
\qquad \mathbf G_\pm(0,z)=I,
\label{eq:flow-family}
\end{equation}
where $\mathbf G_-$ is exterior and normalized by $\mathbf G_-(s,\infty)=I$, while
$\mathbf G_+$ is interior. At $s=2R$,
\begin{equation}
\mathbf G_-(2R,z)=\boldsymbol\Gamma_-(z;R),
\qquad
\mathbf G_+(2R,z)=\boldsymbol\Gamma_+(z;R)^{-1}.
\label{eq:flow-endpoint}
\end{equation}

Differentiate Eq.~\eqref{eq:flow-family}, multiply by $\mathbf G_-^{-1}$ from
the left and by $\mathbf G_+^{-1}$ from the right, and define
\begin{equation}
\widetilde{\mathbf h}(s,z)=\mathbf G_-(s,z)^{-1}\mathbf h(z)\mathbf G_-(s,z).
\label{eq:flow-deformed-h}
\end{equation}
This gives
\begin{equation}
\widetilde{\mathbf h}
=\mathbf G_-^{-1}\partial_s\mathbf G_-+(\partial_s\mathbf G_+)\mathbf G_+^{-1}.
\label{eq:flow-splitting-identity}
\end{equation}
Applying $\mathsf P_-$ and $\mathsf P_+$ yields
\begin{equation}
\boxed{
\partial_s\mathbf G_-=\mathbf G_-\mathsf P_-\widetilde{\mathbf h},
\qquad
\partial_s\mathbf G_+=\mathsf P_+\widetilde{\mathbf h}\,\mathbf G_+ .
}
\label{eq:flow-factor-odes}
\end{equation}
These equations, together with $\mathbf G_\pm(0)=I$, determine the factors.

Differentiating Eq.~\eqref{eq:flow-deformed-h} gives the isospectral
Lax equation
\begin{equation}
\partial_s\widetilde{\mathbf h}
=[\widetilde{\mathbf h},\mathsf P_-\widetilde{\mathbf h}]
=[\mathsf P_+\widetilde{\mathbf h},\widetilde{\mathbf h}].
\label{eq:flow-lax}
\end{equation}
Isospectrality is also immediate from
Eq.~\eqref{eq:flow-deformed-h}: $\widetilde{\mathbf h}$ is conjugate to $\mathbf h$ and
therefore evolves on the fixed elliptic curve
\eqref{eq:spectral-curve-WH}. Moreover,
$\mathbf G_+=\mathbf G_-^{-1}e^{s\mathbf h}$ and $[e^{s\mathbf h},\mathbf h]=0$ imply
\begin{equation}
\widetilde{\mathbf h}=\mathbf G_+\mathbf h\mathbf G_+^{-1},
\label{eq:flow-two-conjugations}
\end{equation}
so the two factors describe the same moving eigenlines.

Equations~\eqref{eq:flow-factor-odes} may be written as ordered
exponentials. With later values of $s$ ordered to the right in
$\overline{\mathcal T}$,
\begin{align}
\boldsymbol\Gamma_-(z;R)
&=\overline{\mathcal T}\exp\!\left[
\int_0^{2R}\mathsf P_-\widetilde{\mathbf h}(s,z)\,ds\right],
\nonumber\\
\boldsymbol\Gamma_+(z;R)
&=\overline{\mathcal T}\exp\!\left[-
\int_0^{2R}\mathsf P_+\widetilde{\mathbf h}(s,z)\,ds\right].
\label{eq:flow-ordered-exponentials}
\end{align}

\subsection{Scalarization on the moving eigenline}

Fix $P\in\mathcal R$ away from the branch points and choose a meromorphic
eigenvector of the undeformed Hamiltonian,
\begin{equation}
\mathbf h(z)r_0(P)=E(P)r_0(P).
\label{eq:flow-eigenvectors}
\end{equation}
At $s=2R$, Eqs.~\eqref{eq:flow-endpoint} and
\eqref{eq:flow-two-conjugations} give
\begin{align*}
\widetilde{\mathbf h}(2R,z)
&=\boldsymbol\Gamma_-(z;R)^{-1}\mathbf h(z)\boldsymbol\Gamma_-(z;R)\\
&=\boldsymbol\Gamma_+(z;R)^{-1}\mathbf h(z)\boldsymbol\Gamma_+(z;R).
\end{align*}
Choose a meromorphic eigenvector $r(P;R)$ of this deformed Hamiltonian,
\[
\widetilde{\mathbf h}(2R,z)r(P;R)=E(P)r(P;R).
\]
Both factors $\boldsymbol\Gamma_\pm$ therefore map the moving eigenline
back to the original eigenline spanned by $r_0(P)$. Define their scalar
amplitudes by
\begin{equation}
\boldsymbol\Gamma_\pm(z;R)r(P;R)
=\chi_\pm(P;R)r_0(P).
\label{eq:flow-scalar-amplitudes}
\end{equation}
Applying $\boldsymbol\Gamma_+=e^{-2R\mathbf h}\boldsymbol\Gamma_-$ to
$r(P;R)$ now gives
\begin{equation}
\chi_+(P;R)=e^{-2RE(P)}\chi_-(P;R),
\qquad P\in\WT.
\label{eq:flow-scalar-jump}
\end{equation}
This is precisely the scalar jump \eqref{eq:scalar-WH-jump}, with
$+$ and $-$ referring to the interior and exterior domains, respectively.

The choice of the meromorphic eigenvector fixes the common scalar
freedom: multiplying $r(P;R)$ by a scalar function multiplies both
$\chi_+$ and $\chi_-$ by that function without changing their jump.
To match the explicit construction in the main text, choose
\[
r(P;R)
=\widetilde{\boldsymbol\Gamma}_-(\infty;R)
\mathbf C(z;R)^{-1}r_0(P).
\]
Substitution into Eqs.~\eqref{eq:preliminary-matrix-WH} and
\eqref{eq:exact-dual-WH-factors} gives
$\chi_\pm(P;R)=\widetilde\chi_\pm(P;R)$ as defined in
Eq.~\eqref{eq:scalar-WH-solution}. Thus the scalar functions agree with
those of the main text, including their auxiliary pole and zero and
their normalization.

The evolution of the eigenline can also be read directly from the flow.
The transported vector
$r_{\mathrm{tr}}(s,P)=\mathbf G_-(s,z)^{-1}r_0(P)$ satisfies
\begin{equation}
\partial_s r_{\mathrm{tr}}(s,P)
=-\mathsf P_-\widetilde{\mathbf h}(s,z)r_{\mathrm{tr}}(s,P),
\qquad r_{\mathrm{tr}}(0,P)=r_0(P).
\label{eq:flow-eigenline-transport}
\end{equation}
At $s=2R$, this vector spans the same eigenline as $r(P;R)$, but
generally has a different scalar normalization. Scalarization is
therefore not a simultaneous diagonalization of the fixed projectors
and the Wiener--Hopf factors: it relates the moving eigenlines to the
original ones.

The factorization flow thus supplies an alternative derivation of the
scalar jump and the evolution of its eigenlines. The theta quotient and
rational correction of Section~\ref{sec:WienerHopf} fix the meromorphic
data needed to reconstruct the complete matrix factors.

\section{Factor swapping and the relation to periodic dimers}
\label{app:alternative-scalar-reduction-factor-swapping}

The main text reduces the canonical factorization of
\(\mathbf W_R(z)=e^{-2R\mathbf h(z)}\) to a scalar Riemann--Hilbert problem by using its
Chebotarev--Khrapkov structure. Here we obtain the same scalar reduction
by repeated factor swapping. For periodic dimer models, the iterative
refactorization was developed in Ref.~\cite{berggren2019correlation},
and its interpretation as the transport of eigenvectors on an invariant
spectral curve was made explicit in
Refs.~\cite{Borodin2022Biased,BerggrenBorodin2023}. In the present
\(2\times2\) problem the eigenlines are one-dimensional, so this transport
reduces to scalar multiplication along each eigenline. The construction
also explains why the relevant eigenlines move during the factorization.
It does not independently solve the global divisor problem: that
information is supplied by the explicit elliptic construction in
Section~\ref{sec:WienerHopf}. Related matrix-orthogonal-polynomial and
matrix Riemann--Hilbert formulations appear in
Refs.~\cite{duits2020two,KuijlaarsPiorkowski2024}.

\subsection{Discrete approximation and factor swapping}

Fix \(R>0\), set \(\epsilon=R/M\), and introduce the one-step symbol
\begin{equation}
\mathbf L_\epsilon(z)=(I-\epsilon \mathbf h(z))(I+\epsilon \mathbf h(z))^{-1}.
\label{eq:swap-one-step}
\end{equation}
For sufficiently large \(M\), this symbol is Hermitian positive definite
on the unit circle: an eigenvalue \(E\in\mathbb R\) of \(\mathbf h(e^{ik})\)
is mapped to \((1-\epsilon E)/(1+\epsilon E)>0\). Moreover,
\begin{equation}
\mathbf L_\epsilon^M
=\exp[-2M\operatorname{arctanh}(\epsilon \mathbf h)]
=e^{-2R\mathbf h+\mathcal O(M^{-2})},
\label{eq:swap-discrete-limit}
\end{equation}
uniformly on \(|z|=1\).

Set \(\boldsymbol\Phi_0=\mathbf L_\epsilon\). At the \(j\)-th step choose the factor order
to agree with the convention of the main text,
\begin{equation}
\boldsymbol\Phi_j=\boldsymbol\Phi_{j,+}\boldsymbol\Phi_{j,-},
\qquad
\boldsymbol\Phi_{j+1}=\boldsymbol\Phi_{j,-}\boldsymbol\Phi_{j,+},
\label{eq:swap-iteration}
\end{equation}
where the interior and exterior factors and their inverses are holomorphic
in their respective regions. Since \(\boldsymbol\Phi_j\) is
positive definite, the factors can be chosen on the unit circle with
\(\boldsymbol\Phi_{j,-}=\boldsymbol\Phi_{j,+}^\dagger\). The swapped symbol
\(\boldsymbol\Phi_{j+1}=\boldsymbol\Phi_{j,+}^\dagger\boldsymbol\Phi_{j,+}\) is therefore positive
definite as well, so the procedure can be iterated. It is also
isospectral, since
\[
\boldsymbol\Phi_{j+1}=\boldsymbol\Phi_{j,+}^{-1}\boldsymbol\Phi_j\boldsymbol\Phi_{j,+}.
\]

Repeatedly using
\((\boldsymbol\Phi_{j,+}\boldsymbol\Phi_{j,-})^m
=\boldsymbol\Phi_{j,+}(\boldsymbol\Phi_{j,-}\boldsymbol\Phi_{j,+})^{m-1}\boldsymbol\Phi_{j,-}\)
gives
\begin{equation}
\mathbf L_\epsilon^M
=\underbrace{\boldsymbol\Phi_{0,+}\boldsymbol\Phi_{1,+}\cdots\boldsymbol\Phi_{M-1,+}}
_{\displaystyle \boldsymbol\Gamma_+^{(M)}}
\underbrace{\boldsymbol\Phi_{M-1,-}\cdots\boldsymbol\Phi_{1,-}\boldsymbol\Phi_{0,-}}
_{\displaystyle (\boldsymbol\Gamma_-^{(M)})^{-1}}.
\label{eq:swap-accumulated-factorization}
\end{equation}
Thus factor swapping directly produces the interior--exterior ordering
\(\mathbf L_\epsilon^M=\boldsymbol\Gamma_+^{(M)}(\boldsymbol\Gamma_-^{(M)})^{-1}\). The usual constant
freedom is fixed by imposing \(\boldsymbol\Gamma_-^{(M)}(\infty)=I\).

\subsection{Scalar transport on the spectral curve}

On the spectral curve \(\mathcal R\) of
Eq.~\eqref{eq:spectral-curve-WH}, the two eigenvalues of
\(\mathbf L_\epsilon\) are represented by the single meromorphic function
\begin{equation}
\lambda_\epsilon(P)
=\frac{1-\epsilon E(P)}{1+\epsilon E(P)},
\qquad
\lambda_\epsilon(P^*)=\lambda_\epsilon(P)^{-1}.
\label{eq:swap-eigenvalue}
\end{equation}
Let \(r_j(P)\) span the \(\lambda_\epsilon(P)\)-eigenline of
\(\boldsymbol\Phi_j\). Away from the branch points this eigenspace is
one-dimensional. Equation~\eqref{eq:swap-iteration} then implies that
there are scalar meromorphic functions \(\alpha_j,\beta_j\) such that
\begin{equation}
\boldsymbol\Phi_{j,-}r_j(P)=\beta_j(P)r_{j+1}(P),
\qquad
\boldsymbol\Phi_{j,+}r_{j+1}(P)=\alpha_j(P)r_j(P).
\label{eq:swap-eigenline-transport}
\end{equation}
Applying the second relation after the first gives
\begin{equation}
\alpha_j(P)\beta_j(P)=\lambda_\epsilon(P).
\label{eq:swap-scalar-one-step}
\end{equation}

Define
\[
A_M(P)=\prod_{j=0}^{M-1}\alpha_j(P),
\qquad
B_M(P)=\prod_{j=0}^{M-1}\beta_j(P).
\]
The accumulated factors in
Eq.~\eqref{eq:swap-accumulated-factorization} act on the transported
eigenlines as
\begin{equation}
\boldsymbol\Gamma_+^{(M)}r_M(P)=A_M(P)r_0(P),
\qquad
(\boldsymbol\Gamma_-^{(M)})^{-1}r_0(P)=B_M(P)r_M(P),
\label{eq:swap-accumulated-transport}
\end{equation}
and Eq.~\eqref{eq:swap-scalar-one-step} gives
\begin{equation}
A_M(P)B_M(P)=\lambda_\epsilon(P)^M.
\label{eq:swap-scalar-product}
\end{equation}
Consequently, with
\begin{equation}
\chi_+^{(M)}(P)=A_M(P),
\qquad
\chi_-^{(M)}(P)=B_M(P)^{-1},
\label{eq:swap-scalar-factors}
\end{equation}
the matrix factorization reduces to the scalar jump
\begin{equation}
\boxed{
\chi_+^{(M)}(P)
=\lambda_\epsilon(P)^M\chi_-^{(M)}(P).
}
\label{eq:swap-scalar-jump}
\end{equation}

For completeness, introduce the meromorphic eigenvector frame
\[
\mathbf R_j(z)=\begin{pmatrix}r_j(P)&r_j(P^*)\end{pmatrix}.
\]
Away from the branch points the matrix factors are reconstructed as
\begin{align}
\boldsymbol\Gamma_+^{(M)}
&=\mathbf R_0
\operatorname{diag}\bigl(A_M(P),A_M(P^*)\bigr)\mathbf R_M^{-1},
\nonumber\\
(\boldsymbol\Gamma_-^{(M)})^{-1}
&=\mathbf R_M
\operatorname{diag}\bigl(B_M(P),B_M(P^*)\bigr)\mathbf R_0^{-1}.
\label{eq:swap-matrix-reconstruction}
\end{align}
Apparent singularities of the frames at the branch points cancel in the
complete matrix expressions. Rescaling \(r_j\) by a nonzero meromorphic
function multiplies \(\chi_+^{(M)}\) and \(\chi_-^{(M)}\) by the same
meromorphic factor and leaves both their ratio and
Eq.~\eqref{eq:swap-matrix-reconstruction} unchanged. This is the divisor
freedom encoded in the main text by the theta quotient and the rank-one
rational correction.

\subsection{Continuous-time limit}

For fixed \(R\), as \(M\to\infty\),
\begin{equation}
M\log\lambda_\epsilon(P)
=-2RE(P)+\mathcal O(M^{-2})
\label{eq:swap-continuous-eigenvalue}
\end{equation}
uniformly on the lifted Wiener--Hopf contour. Hence
Eq.~\eqref{eq:swap-scalar-jump} converges to
\begin{equation}
\boxed{
\chi_+(P)=e^{-2RE(P)}\chi_-(P),
}
\label{eq:swap-continuous-jump}
\end{equation}
which is precisely Eq.~\eqref{eq:scalar-WH-jump}. Stability of normalized
canonical factorizations then gives convergence of the matrix factors;
in the scalar representation one chooses compatible meromorphic frames
so that their divisor data do not cross the contour. Thus factor swapping
derives the scalar jump and the transport of its eigenline, while the
construction in Section~\ref{sec:WienerHopf} supplies the complete global
solution on the elliptic curve.

\section{Hydrodynamic picture}
\label{app:hydrodynamic-picture}

This appendix gives a complementary hydrodynamic interpretation of the
characteristics used in Sections~\ref{sec:main_results} and
\ref{sec:saddle-geometry}. It is not needed for the exact
Wiener--Hopf derivation. Its purpose is to explain why the answer takes the
form of free quasiparticle motion, how the return boundary condition fixes
the effective midpoint curve, and why the two arctic boundaries arise as
caustics of the same characteristic surface. We take $N\to\infty$ first
and then let $R\to\infty$ with $x=m/R$ and $t=\tau/R$ fixed.

\subsection{Free characteristics on the spectral curve}

Let \(\eta(z)\) denote a local reference branch of \(\sqrt{f(z)}\), and write
\[
P_\sigma=(z,\sigma\eta(z)),\qquad \sigma=\pm,
\]
for the corresponding points on the two sheets of the spectral curve
\eqref{eq:spectral-curve-WH}. The band energy and group velocity are
\begin{equation}
E_\sigma(z)=\sigma\frac{\eta(z)}{z},
\qquad
v_\sigma(z)=z\partial_zE_\sigma(z)
=\sigma\frac{z^2-1}{2\eta(z)}.
\label{eq:hydro-velocity}
\end{equation}
Writing the local complex quasimomentum as
$p_\sigma(x,t)=\log z_\sigma(x,t)$, the Euler-scale free-fermion equation
takes the complex Burgers form used in the hydrodynamic description of the
emptiness formation probability \cite{Abanov2006Hydrodynamics} and, more
generally, in limit-shape and inhomogeneous-fermion hydrodynamics
\cite{Kenyon2007,CFTstephan,bocini2021non}:
\begin{equation}
\partial_t p_\sigma
+v_\sigma(e^{p_\sigma})\partial_xp_\sigma=0.
\label{eq:hydro-equation}
\end{equation}
The spectral parameter is therefore constant along a characteristic, and
the general solution is
\begin{equation}
x_\sigma(t,z)=x_{0,\sigma}(z)+t\,v_\sigma(z).
\label{eq:hydro-free-motion}
\end{equation}
Free propagation determines the second term, but it does not determine the
midpoint curve $x_{0,\sigma}(z)$. The latter contains the information about
the Euclidean return boundary condition.

\subsection{Midpoint data from the domain wall}

It is useful to encode the midpoint curve in the differential
\begin{equation}
d\Omega_{\mathrm{mid},\sigma}
=x_{0,\sigma}(z)\,d\log z.
\label{eq:hydro-midpoint-definition}
\end{equation}
The domain-wall state singles out the two punctures $z=0$ and $z=\infty$,
which represent the two limits of infinite complex momentum. The
differential must have the hard-edge principal parts required at these
punctures, but no finite poles: an additional finite pole would describe a
boundary defect or a momentum-selective source absent from the problem.
These conditions determine the differential up to a multiple of the
holomorphic differential $dz/\eta$. The remaining ambiguity is removed by
requiring zero period around the compact $a$-cycle. Physically, a nonzero
period would introduce an additional exponential monodromy around the
internal band cycle, whereas the domain-wall state carries no such twist.

The unique normalized differential with these properties is
\begin{equation}
d\Omega_{\mathrm{mid},\sigma}
=\sigma\frac{z^2-2cz+1}{2z\eta(z)}\,dz,
\qquad
c=\frac{\displaystyle\oint_a z\,dz/\eta}
        {\displaystyle\oint_a dz/\eta},
\label{eq:hydro-midpoint-differential}
\end{equation}
where $c$ is the same constant as in Eq.~\eqref{eq:c-WH}. Hence
\begin{equation}
x_{0,\sigma}(z)
=\sigma\frac{z^2-2cz+1}{2\eta(z)},
\qquad
x_\sigma(t,z)
=\sigma\frac{(1+t)z^2-2cz+(1-t)}{2\eta(z)}.
\label{eq:hydro-characteristics}
\end{equation}
The inversion $z\mapsto z^{-1}$ leaves $x_{0,\sigma}$ invariant and
reverses $v_\sigma$. It therefore exchanges the temporal boundaries
$t=-1$ and $t=1$, as required by the return geometry.

The same midpoint differential follows directly from the exact
Wiener--Hopf solution. Indeed, the second-kind differentials
$d\Omega_0$ and $d\Omega_\infty$ defined in
Eq.~\eqref{eq:second-kind-WH} satisfy
\begin{equation}
d\Omega_{\mathrm{mid},\sigma}
=\frac{\sigma}{2}\left(d\Omega_0-d\Omega_\infty\right).
\label{eq:hydro-WH-identification}
\end{equation}
Thus the hydrodynamic prescription is the Euler-scale content of the
canonical factorization. The theta quotient and the rational divisor
correction are essential for the exact factors, but they contain no
$z$-dependent exponential proportional to $R$ and do not change the
characteristic surface. Notice also that the same normalization constant
$c$ fixes the leading return free energy in
Section~\ref{sec:return amplitude}.

\subsection{Caustics and local phases}

The projection of the characteristic surface
\eqref{eq:hydro-characteristics} to the $(x,t)$ plane becomes singular
when
\begin{equation}
\partial_zx_\sigma(t,z)=0.
\label{eq:hydro-caustic}
\end{equation}
This is precisely the double-saddle condition
\eqref{eq:arctic-double-root}. Its positive-real component gives the outer
frozen--liquid curve, while the compact interval
$(-\kappa,-\kappa^{-1})$ gives the inner liquid--gas curve. Evaluating
Eq.~\eqref{eq:hydro-caustic} produces the parametrization
\eqref{eq:results-arctic-curves}; the detailed classification of its roots
is given in Section~\ref{sec:saddle-geometry}.

Hydrodynamics fixes the characteristic geometry but not, by itself, the
complete microscopic state. The occupied spectral arcs and the residue of
the Cauchy pole in the exact kernel determine that state, as explained in
Section~\ref{sec:transition-kernel} and
Appendix~\ref{app:steepest-descent-topology}. In the liquid the occupied
arc has moving endpoints and produces algebraic correlations. Inside the
inner caustic it closes into the full lower-band cycle, producing the
translation-invariant, half-filled, exponentially correlated gas state.
In this sense the hydrodynamic caustic and the Wiener--Hopf contour
deformation give complementary macroscopic and microscopic descriptions
of the same phase boundaries.

\section{Extension to the Rice--Mele chain}
\label{app:rice-mele-extensions}

This appendix extends the leading return and arctic geometry of the
staggered chain studied in the main text to the full Rice--Mele model,
allowing both unequal intracell and intercell hoppings and a staggered
on-site potential. The dispersion reduces to that of the uniform-hopping
model after introducing an effective hopping scale and an effective gap.
Consequently, the spectral curve, leading return free energy, and
Euler-scale arctic boundaries retain the same form under a simple
reparametrization.

Using the unit-cell convention of Section~\ref{sec:model}, let the
intracell and intercell hoppings be $v$ and $w$, respectively, with
$v,w>0$. For the Rice--Mele Hamiltonian
\eqref{eq:rice-mele-hamiltonian}, the Bloch matrix is
\begin{align}
\mathbf h_{\rm RM}(k)
&=\begin{pmatrix}
\Delta & v+w e^{-ik}\\
 v+w e^{ik} & -\Delta
\end{pmatrix}
=\bm d(k)\cdot\bm\sigma,
\label{eq:app-rm-bloch}\\
\bm d(k)
&=\left(v+w\cos k,\,w\sin k,\,\Delta\right).
\label{eq:app-rm-d-vector}
\end{align}
The two bands have energies $E_\sigma(k)=\sigma\varepsilon(k)$,
$\sigma=\pm$, with
\begin{equation}
\varepsilon(k)^2
=v^2+w^2+\Delta^2+2vw\cos k.
\label{eq:app-rm-dispersion-raw}
\end{equation}
The bulk gap closes only at $v=w$ and $\Delta=0$. It is useful to define
\begin{equation}
J\coloneqq\sqrt{vw}>0,
\qquad
\mu^2\coloneqq\Delta^2+(v-w)^2,
\label{eq:app-rm-J-mu}
\end{equation}
so that
\begin{equation}
\varepsilon(k)^2=\mu^2+4J^2\cos^2\frac{k}{2}.
\label{eq:app-rm-dispersion}
\end{equation}
Thus the dispersion matches that of the model studied in the main text,
with hopping scale \(J\) and effective gap \(\mu\). Moreover, since
\(\mathbf h_{\rm RM}(z)\) is traceless and satisfies
\[
\mathbf h_{\rm RM}(z)^2=\varepsilon(z)^2I,
\]
the symbol retains the same Chebotarev--Khrapkov scalar-square structure.
The eigenvector frame changes, and therefore so do the matrix prefactors
and sublattice-resolved observables, but the Abelian exponential
controlling the leading return free energy and the saddle action depends
only on the spectral curve. Consequently, the leading scalar geometry
carries over under the reparametrization below.

We therefore introduce
\begin{equation}
A=2+\frac{\mu^2}{J^2},
\qquad
\kappa=\frac{A+\sqrt{A^2-4}}{2},
\qquad
\kappa+\kappa^{-1}=A,
\label{eq:app-rm-A-kappa}
\end{equation}
and
\begin{equation}
c=-\frac{1}{\kappa}
\frac{\mathrm E(1-\kappa^2)}{\mathrm K(1-\kappa^2)}.
\label{eq:app-rm-c}
\end{equation}
The corresponding spectral curve is
\begin{equation}
\mathcal R_{\rm RM}:\qquad
\eta^2
=J^2 z(z+\kappa)(z+\kappa^{-1})
\equiv f(z).
\label{eq:app-rm-spectral-curve}
\end{equation}
On the physical unit circle $z=e^{ik}$ we choose the branch
$\eta=z\varepsilon(k)$. With this rescaling, the spectral-curve and
saddle-point analysis that determines the leading return free energy and
the arctic geometry carries over from the uniform-hopping model. In
particular,
\begin{equation}
\log Z(R)=-4J^2cR^2+\mathcal O(1),
\label{eq:app-rm-return}
\end{equation}
and the two components of the arctic boundary are generated by the same
real spectral intervals as in the main text, with parametrization
\begin{align}
t(z)
&=-\frac{(z^2-1)\left[z^2+2(A+c)z+1\right]}
{z^4+2Az^3+6z^2+2Az+1},
\label{eq:app-rm-arctic-t}\\
x_\sigma(z)
&=\sigma\,
\frac{2\eta(z)\left[2z-c(z^2+1)\right]}
{z^4+2Az^3+6z^2+2Az+1},
\qquad \sigma=\pm.
\label{eq:app-rm-arctic-x}
\end{align}
The positive interval $z\in(0,\infty)$ gives the outer frozen--liquid
boundary, while $z\in(-\kappa,-\kappa^{-1})$ gives the inner
liquid--gas boundary. The uniform-hopping formulas of the main text are
recovered by setting $v=w=J=1$, for which $\mu^2=\Delta^2$.

The SSH chain is obtained by setting $\Delta=0$. In this limit,
\begin{equation}
J=\sqrt{vw},
\qquad
\mu=|v-w|,
\qquad
A=\frac{v}{w}+\frac{w}{v},
\qquad
\kappa=\max\left(\frac{v}{w},\frac{w}{v}\right).
\label{eq:app-rm-ssh-parameters}
\end{equation}
The SSH gap closes at $v=w$, where $\mu=0$ and $\kappa=1$. For $v\neq w$,
Eqs.~\eqref{eq:app-rm-return}--\eqref{eq:app-rm-arctic-x} therefore give
the leading return free energy and arctic boundaries of the gapped SSH
chain. At fixed $J$, these quantities depend on the dimerization $v-w$
only through $(v-w)^2$. Consequently, the topological and trivial SSH
dimerizations have the same leading scalar arctic geometry, although
sublattice-resolved observables and boundary or interface physics can
distinguish them.

\bibliography{arctic-gas-linked}

\nolinenumbers

\end{document}